# Classification of Symmetry Derived Pairing at M Point in FeSe

P. Myles Eugenio and Oskar Vafek
*Florida State University and the National High Magnetic Field Lab*

Using the constraints imposed by the crystalline symmetry of FeSe and the experimentally observed phenomenology, we analyze the possible pairing symmetry of the superconducting order parameter focusing on intercalated and monolayer FeSe compounds. Such analysis leads to three possible pairing symmetry states – s-wave, d-wave, and helical p-wave. Despite the differences in the pairing symmetry, each of these states is fully gapped with gap minimum centered above the normal state Fermi surface, in agreement with photoemission data of Y. Zhang *et al.* [7] The analysis provides additional insights into the possible pairing mechanism for each of these states, highlighting the detrimental role of the renormalized repulsive intra-orbital Hubbard $U$ and inter-orbital $U'$, and the beneficial role of the pair hopping $J'$ and the Hunds $J$ terms, as well as the spin-orbit coupling in the effective low-energy Hamiltonian.

## I. INTRODUCTION

The wealth of physical phenomena exhibited by the iron-based superconductors has lead to an active field of research with challenging open questions [1, 2]. Notable among them is the pairing symmetry and the mechanism of high temperature superconductivity which they exhibit [3, 4]. Recently the focus has shifted towards the iron-selenide (FeSe) family of superconductors, with reported transition temperatures as high as 8K in the bulk [5], 40 K in ($Li_{1-x}Fe_x$)OHFeSe [6], 65 K in monolayer FeSe grown on a $SrTiO_3$ [7], and even 109 K [8] in the latter system.

The monolayer FeSe is fundamentally a single iron plane with selenium atoms puckered in and out of that plane Fig(1). In practice the 2D plane is grown on a substrate (e.g. $SrTiO_3$), leading to undoped [11] or doped [7, 9, 10] monolayer FeSe. Bulk FeSe [12–14] is a three dimensional crystal composed of vertically aligned FeSe planes. Additional three dimensional arrangements can be formed by sandwiching intercalates between the FeSe planes [6, 15–17], where each stack is connected by a weak inter-layer coupling [17].

The family of FeSe superconductors can thus be viewed as different arrangements of the *same material*, rather than altogether different materials. This suggests that the common structural unit, namely the FeSe plane, is responsible for the common electronic properties and, importantly, for superconductivity. Any differences in the physical characteristics – such as differences in $T_c$ or the appearance of nematicity in the bulk FeSe, etc – then presumably arise from differences in doping, strength of the interlayer coupling, inversion symmetry breaking of monolayer on a substrate, or from a non-electronic origin (e.g. interface phonons) [9, 10].

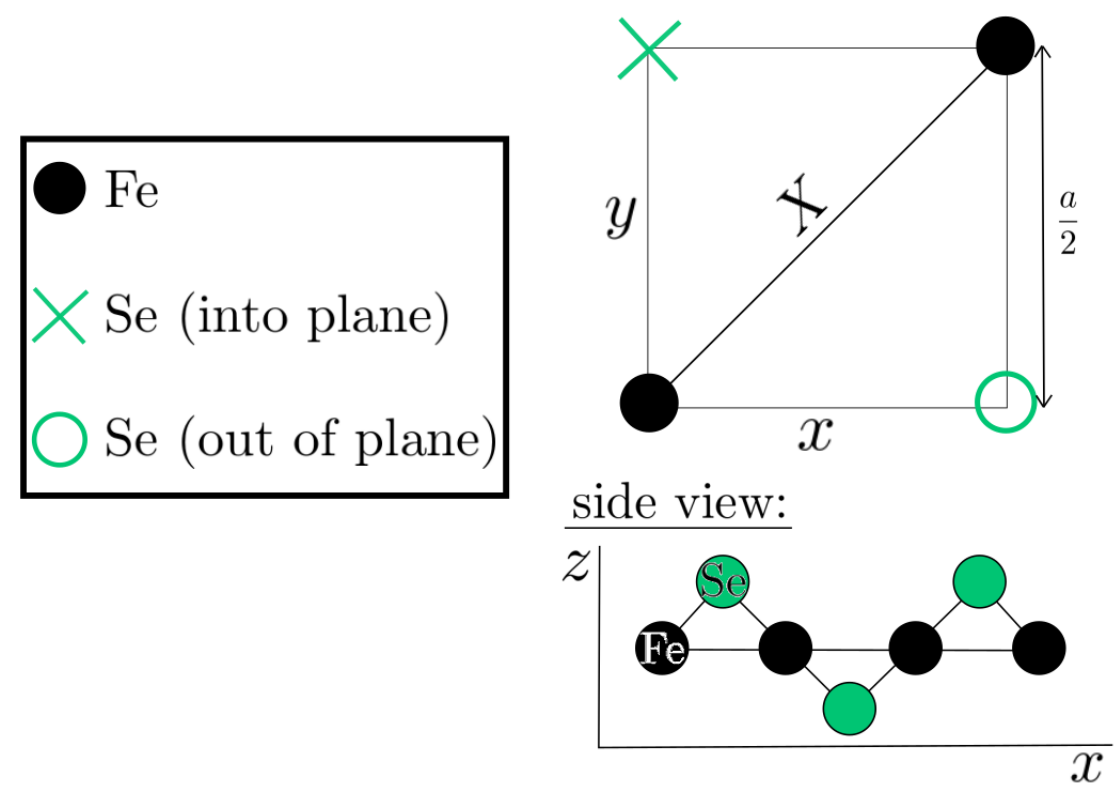

FIG. 1. The unit cell of the FeSe plane. There are 2-Fe per unit cell, with Se-atoms puckered above and below the Fe-plane. The constant $a$ is the lattice spacing. The space group of the iron-selenide is discussed in Ref[19]. The three space group generators [19] are a mirror across the yz-plane $m_x$, a mirror across the Yz-plane $m_X$ followed by a $(\frac{a}{2}, \frac{a}{2})$ translation, and a xy-plane mirror $m_z$ followed by a $(\frac{a}{2}, \frac{a}{2})$ translation. Where $X = x + y$ and $Y = -x + y$.

Indeed the common electronic feature to these materials is the presence of two electron-like Fermi surfaces centered at the Brillouin zone edge (M-point). Further, Angularly Resolved Photo-Emission Spectroscopy (ARPES) in the bulk [12, 14], intercalated [16, 17], and monolayer [7, 9] shows that the electron-like Fermi surface bands originate from two separate binding energies at the M-point (see Fig 2). The bulk FeSe further exhibits a hole Fermi surface at the Brillouin zone center (Γ-point) [12, 14]; the intercalated and monolayer systems have only electron Fermi surfaces. These differences can be understood to be primarily due to the differences in the Fermi energy (doping), as opposed to changes in the bandstructure.

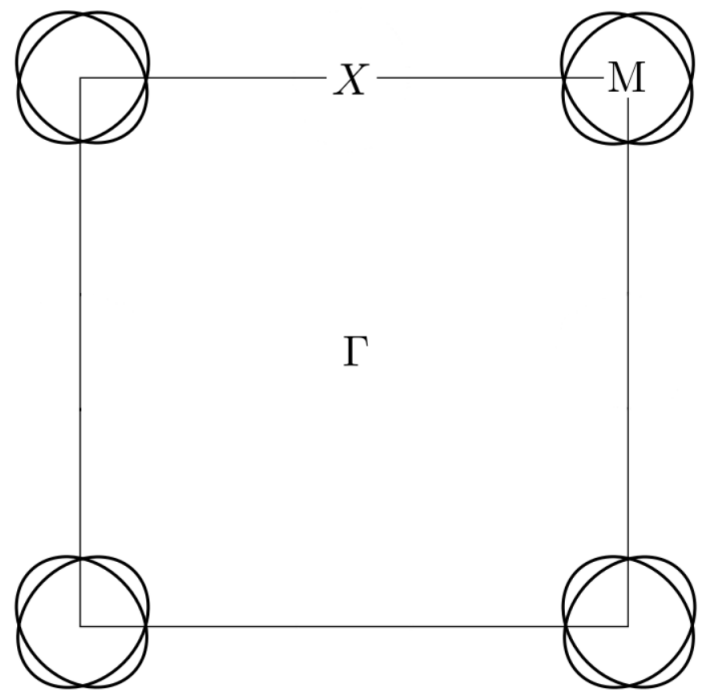

FIG. 2. Sketch of Fermi surfaces as seen in intercalated and monolayer FeSe. There are two points of high symmetry: the Brillouin zone center $\Gamma = (0, 0)$ and the zone edge $\mathrm{M} = (\frac{\pi}{a}, \frac{\pi}{a})$. In the proper 2-Fe/UC picture (shown here), two electron pockets cross at the M-point. The Fermi surface crossing occurs along the M$X$-direction, where $X$ marks the Brillouin zone boundary.

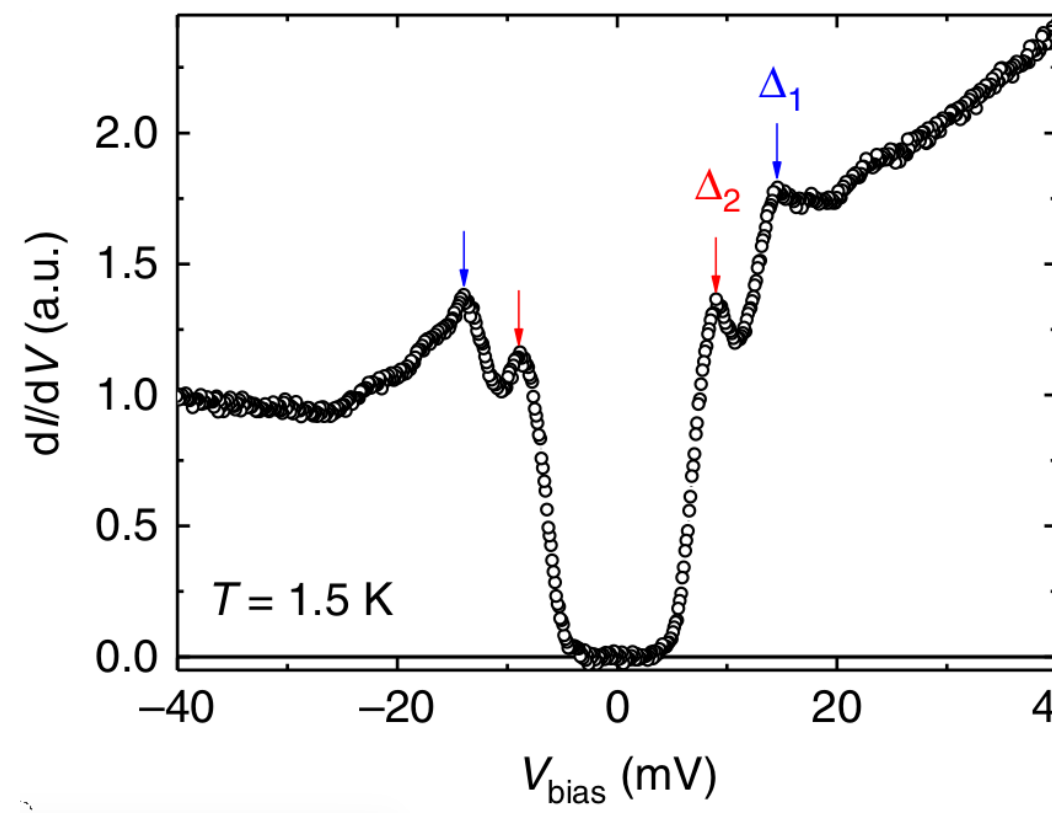

FIG. 3. Experimental result Fig(2.C) taken from Ref[6] under the Creative Commons Attribution 4.0 International License.

The superconducting gap in $(\mathrm{Li}_{1-x}\mathrm{Fe}_x)$OHFeSe is reported to be nodeless, nearly isotropic with gap size $(13 \pm 2)$ meV in Ref[17] and $\sim$ 10 meV in Ref[16]. Further, synchrotron ARPES in the monolayer shows an anisotropic gap, varying from 8 meV to 13 meV [7]. The superconducting gap is also claimed to show "back-bending", i.e. the gap sits directly above the normal state Fermi surfaces [7]. This last observation, upon which we elaborate later, is not trivial in that the separation between the two bands that cross the Fermi surface in the ΓM-direction is only about $\sim 15$ meV (see Fig 2 in Ref[7]), thus comparable to the pairing gap itself.

Scanning Tunnelling Microscopy (STM) experiments also support the existence of a large superconducting gap in these materials [6, 18]. More interestingly, they show a hard gap followed by not one but *two* peaks in the $dI/dV$, Fig(3) [6]. The two peaks occur at 8.6 meV and 14.3 meV in intercalated $(\mathrm{Li}_{1-x}\mathrm{Fe}_x)$OHFeSe [6], and 9 meV and 20.1 meV in monolayer[18]. The suggestion that the higher energy peak directly reflects the superconducting gap is at odds with the monolayer ARPES data, whose gap maximum is, as we stated, ∼13 meV (as shown in Fig 4 of Ref [7], it never exceeds 14 meV, even considering experimental uncertainty).

In this paper, we concentrate on understanding to what degree do the crystalline symmetry of FeSe, and the experimentally observed phenomenology mentioned above, constrain the possible pairing symmetry of the superconducting order parameter. We focus on intercalated $(\mathrm{Li}_{1-x}\mathrm{Fe}_x)$OHFeSe and monolayer FeSe epitaxially grown on $\mathrm{SrTiO}_3$ where the 3D dispersion effects are absent (although we ignore the inversion symmetry breaking due to the substrate in the monolayer FeSe). To this end, we build upon the low energy model that respects the full space group symmetry of the material and includes spin-orbit effects [19], but extend it in ways that better account for the observed phenomenology of FeSe. Such analysis leads us to three possible pairing symmetry states – s-wave, d-wave, and helical p-wave. Despite the differences in the pairing symmetry, each of these states is fully gapped with gap minimum centered above the normal state Fermi surface. Our analysis also gives us insight into the possible pairing mechanism for each of these states, highlighting the detrimental role of the repulsive intra-orbital Hubbard $U$ and inter-orbital $U'$, and the beneficial role of the pair hopping $J'$ and the Hunds $J$ terms in the effective low-energy Hamiltonian. (As explained in Sec(II B) each of the interaction couplings should be understood as renormalized and orbital, or more precisely, Bloch function -dependent.) The Hunds coupling, together with spin orbit coupling, was recently proposed by one of us and Chubukov to explain the phenomenology and the mechanism of pairing in $\mathrm{KFe}_2\mathrm{As}_2$ [20]. This is another example of an iron-based superconductor with only one carrier type (hole) Fermi surface.

The paper is organized as follows: In section 2, we introduce the low energy effective model – including a new momentum-dependent *intra-pocket* spin-orbit coupling – as well as the symmetry allowed interaction couplings. In section 3, we determine the values of the symmetry allowed parameters in the normal state based on the detailed comparison with the ARPES experiment. In section 4 we classify all pairing states based on the symmetry. Armed with the values of the parameters obtained in section 3, we then highlight the details of the phenomenology of the superconducting state, and critically compare them with the predictions for all pairing states. Assuming an overall time reversal symmetry, we find that only s-, d- and helical p-states are compatible with the mentioned phenomenology. Section 5 provides detailed analysis of the three pairing states. The discussion and the outlook are presented in the final section.

## II. MODEL

We employ an itinerant model developed in Ref[19] for the electron-like Fermi surfaces at the M-point. This low energy effective theory is constructed by requiring invariance under the FeSe space group symmetries and time reversal. Using the nomenclature of Ref[19], the two electronically relevant M-point representations are M1 and M3. We construct doublets $\psi_{X,\alpha}(\mathbf{k}) = (1_{X,\alpha}(\mathbf{k}), 3_{X,\alpha}(\mathbf{k}))^T$ and $\psi_{Y,\alpha}(\mathbf{k}) = (1_{Y,\alpha}(\mathbf{k}), 3_{Y,\alpha}(\mathbf{k}))^T$. Our starting Hamiltonian in the normal state is:

$$H_0 = \sum_{\mathbf{k}} \sum_{\alpha\beta=\uparrow,\downarrow} \psi^\dagger_{M,\alpha}(\mathbf{k}) \begin{pmatrix} h'^{\alpha\beta}_X(\mathbf{k}) & \Lambda_{\alpha\beta} \\ \Lambda^\dagger_{\alpha\beta} & h'^{\alpha\beta}_Y(\mathbf{k}) \end{pmatrix} \psi_{M,\beta}(\mathbf{k}), \tag{1}$$

where

$$\psi_{M,\alpha}(\mathbf{k}) = \left(\psi_{X,\alpha}(\mathbf{k}), \psi_{Y,\alpha}(\mathbf{k})\right)^T. \tag{2}$$

The matrix $h'_X$ in Eqn(1) is the Hamiltonian for one electron-like pocket:

$$h'^{\alpha\beta}_X(\mathbf{k}) = h_X(\mathbf{k})\delta^{\alpha\beta} + \left(\lambda_z(k_x-k_y) + p_{z1}(k_x^3-k_y^3) + p_{z2}k_xk_y(-k_x+k_y)\right)\sigma_3^{\alpha\beta}\tau_1, \tag{3}$$

where $\tau_i$ and $\sigma_i$ are Pauli-matrices acting in orbital and spin space respectively. This is an extension of the Hamiltonian for a single electron pocket $h_X$, developed in Ref[19]:

$$\begin{aligned} h_X(\mathbf{k}) &= \begin{pmatrix} \epsilon_1 + \frac{k^2}{2m_1} + a_1k_xk_y & -iv(k_x+k_y) \\ iv(k_x+k_y) & \epsilon_3 + \frac{k^2}{2m_3} + a_3k_xk_y \end{pmatrix} \\ &\equiv h_{X0}(\mathbf{k}) + h_{X3}(\mathbf{k})\tau_3 + h_{X2}(\mathbf{k})\tau_2. \end{aligned} \tag{4}$$

Because the Hamiltonian $h'^{\alpha\beta}_X(\mathbf{k})$ is diagonal in spin-space, its spin-diagonal elements will be referred to as $h'^{\uparrow}_X \equiv h'^{\uparrow\uparrow}_X$ and $h'^{\downarrow}_X \equiv h'^{\downarrow\downarrow}_X$.

The motivation for extending the Hamiltonian is explained in Sec(III). The prefactor of the k-linear term is denoted $\lambda_z$, because it couples the orbital degrees of freedom with the out-of-plane spin $\sigma_z$; k-cubic terms were also introduced with prefactors $p_{z1}$ and $p_{z2}$. It should be noted that this momentum-dependent spin-orbit acts *within* each pocket and does not mix the electron pockets, i.e. it is intra-band. The extended Hamiltonian for the second electron pocket at the M-point can again be obtained by performing a mirror reflection in the yz-plane:

$$h'_Y(k_x,k_y) = \frac{\sigma_1-\sigma_2}{\sqrt{2}} h'_X(-k_x,k_y) \frac{\sigma_1-\sigma_2}{\sqrt{2}}. \tag{5}$$

Note that $\sigma_3\tau_1$ changed sign under the mirror reflection, because $\sigma_3$ is an axial vector.

It was further shown in Ref[19] that a momentum-independent inter-band spin-orbit coupling is allowed by symmetry. Such term, parameterized by $\lambda$, comes from the coupling of the orbital degrees of freedom with the in-plane spin-vector $\vec{\sigma} = (\sigma_X, \sigma_Y)$, thus breaking the spin SU(2) symmetry:

$$\begin{aligned} h_{SOC} &= \sum_{\mathbf{k}} \sum_{\alpha\beta=\uparrow,\downarrow} \psi^\dagger_{X,\alpha}\Lambda_{\alpha\beta}\psi_{Y,\beta} + h.c. \\ &= \sum_{\mathbf{k}} \psi^\dagger_{X,\uparrow} \begin{pmatrix} 0 & i\lambda \\ \lambda & 0 \end{pmatrix} \psi_{Y,\downarrow} + \psi^\dagger_{X,\downarrow} \begin{pmatrix} 0 & i\lambda \\ -\lambda & 0 \end{pmatrix} \psi_{Y,\uparrow} + h.c. \end{aligned} \tag{6}$$

Another important consequence of this (inter-band) $h_{SOC}$ is the lifting of the degeneracy in the direction of the Fermi surface crossing, leading to the formation an “inner” and “outer” Fermi surface (see Fig 6).

### A. Bloch Sphere

The Hamiltonian for an electron pocket $h'_X$ has one band that disperses downward and one band that disperse upward and crosses the Fermi level. Because the difference in energy between the bands at the Fermi level and the bands below the Fermi level is an order of magnitude larger than the pairing scale, it is useful to project onto the prior. This reduces the size of Hamiltonian by half, facilitating the symmetry analysis.

This projected basis can be visualized in terms of a Bloch sphere Fig(4). For each spin-diagonal element, we write the non-identity part of Eqn(3) as

$$\begin{aligned} \not{h} &\equiv h'^{\uparrow}_X(\mathbf{k}) - h_{X0}(\mathbf{k}) \\ &= h_{X3}(\mathbf{k})\tau_3 + h_{X2}(\mathbf{k})\tau_2 + h_{X1}(\mathbf{k})\tau_1. \end{aligned} \tag{7}$$

Noting that $\not{h}$ and $h'^{\uparrow}_X$ have equivalent eigenstates. We can then define

$$\hat{\not{h}} \equiv \frac{\not{h}}{|\not{h}|} = \begin{pmatrix} \cos\theta & \sin\theta e^{-i\phi} \\ \sin\theta e^{i\phi} & -\cos\theta \end{pmatrix}, \tag{8}$$

where the Bloch angles are functions of momentum $(\theta(\mathbf{k}), \phi(\mathbf{k}))$. The up- and down-spin Hamiltonians map into each other as $h'^{\uparrow}_X(\theta, \pi-\phi) = h'^{\downarrow}_X(\theta,\phi)$, and the two electron pockets map into each other as $h'^{\uparrow}_X(\theta(-k_x,k_y), \pi - \phi(-k_x,k_y)) =$

$h'^{\uparrow}_{Y}(\theta(k_x,k_y),\phi(k_x,k_y))$.

The eigenstate of $h'^{\uparrow}_{X}$ that crosses the Fermi level has the form

$$|X\uparrow>=\left(\cos\frac{\theta}{2}\,e^{-i\phi},\sin\frac{\theta}{2}\right)^{T}. \quad (9)$$

All other eigenstates can be obtained through the above mentioned symmetry relationships.

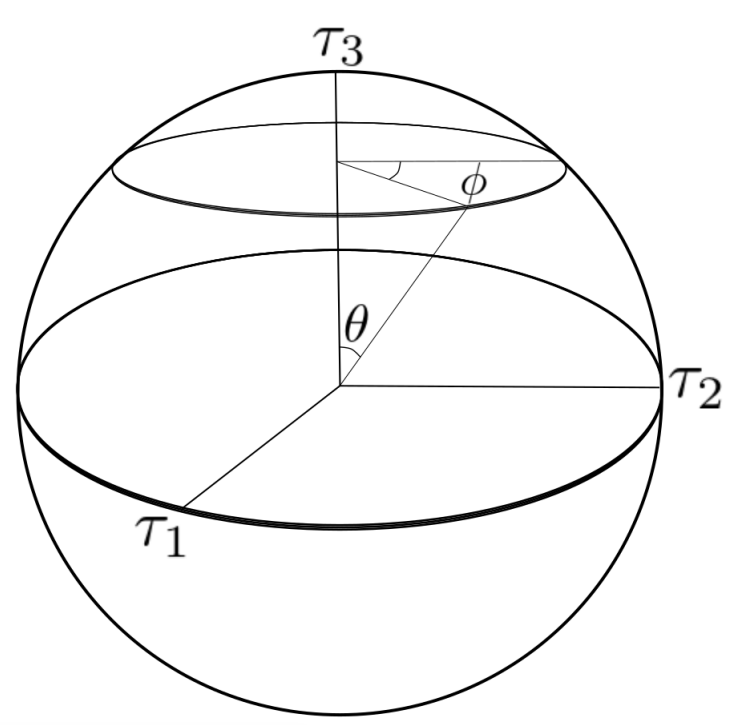

FIG. 4. The Bloch Sphere. The unit sphere embedded in a 3D space, where each orthogonal direction is associated with an SU(2) generator. The north pole ($+\tau_3$ direction) and south pole ($-\tau_3$ direction) represent the M1 and M3 reference states respectively.

### B. Interactions

There are 14 SU(2) invariant couplings at the M-point which contribute to the interacting Hamiltonian $H^{\rm M}_{\rm int}$. We write each interaction in terms of the M1 and M3 symmetry representations, using the doublets $1_\alpha=(1_{X,\alpha},1_{Y,\alpha})^T$ and $3_\alpha=(3_{X,\alpha},3_{Y,\alpha})^T$. The 14 couplings are listed in Eqn(10) in a form facilitating to applying mean field, and with each coupling constant being labelled by the symmetry of the corresponding mean field order parameter listed in Table III,IV, and V.

$$\begin{aligned}
H^{\rm M}_{\rm int}=&\sum_{\mathbf{k}}\sum_{\alpha,\beta\in\{\uparrow,\downarrow\}} g^{(1)}_{A_{1g}}1^\dagger_\alpha(\mathbf{k})\tau_0 1^*_\beta(-\mathbf{k})1^T_\beta(-\mathbf{k})\tau_0 1_\alpha(\mathbf{k})\\
&+g^{(1)}_{B_{2g}}1^\dagger_\alpha(\mathbf{k})\tau_3 1^*_\beta(-\mathbf{k})1^T_\beta(-\mathbf{k})\tau_3 1_\alpha(\mathbf{k})\\
&+g^{(1)}_{A_{2u}}1^\dagger_\alpha(\mathbf{k})\tau_1 1^*_\beta(-\mathbf{k})1^T_\beta(-\mathbf{k})\tau_1 1_\alpha(\mathbf{k})\\
&+g^{(1)}_{B_{1u}}1^\dagger_\alpha(\mathbf{k})\tau_2 1^*_\beta(-\mathbf{k})1^T_\beta(-\mathbf{k})\tau_2 1_\alpha(\mathbf{k})\\
&+g^{(3)}_{A_{1g}}3^\dagger_\alpha(\mathbf{k})\tau_0 3^*_\beta(-\mathbf{k})3^T_\beta(-\mathbf{k})\tau_0 3_\alpha(\mathbf{k})\\
&+g^{(3)}_{B_{2g}}3^\dagger_\alpha(\mathbf{k})\tau_3 3^*_\beta(-\mathbf{k})3^T_\beta(-\mathbf{k})\tau_3 3_\alpha(\mathbf{k})\\
&+g^{(3)}_{B_{2u}}3^\dagger_\alpha(\mathbf{k})\tau_1 3^*_\beta(-\mathbf{k})3^T_\beta(-\mathbf{k})\tau_1 3_\alpha(\mathbf{k})\\
&+g^{(3)}_{A_{1u}}3^\dagger_\alpha(\mathbf{k})\tau_2 3^*_\beta(-\mathbf{k})3^T_\beta(-\mathbf{k})\tau_2 3_\alpha(\mathbf{k})\\
&+g^{t}_{E_u}\Big(\big(1^\dagger_\alpha(\mathbf{k})\tau_0(\sigma_2\sigma_j)_{\alpha\beta}3^*_\beta(-\mathbf{k})\big)\big(3^T_\mu(-\mathbf{k})\tau_0(\sigma_j\sigma_2)_{\mu\nu}1_\nu(\mathbf{k})\big)\\
&\qquad+\big(1^\dagger_\alpha(\mathbf{k})\tau_3(\sigma_2\sigma_j)_{\alpha\beta}3^*_\beta(-\mathbf{k})\big)\big(3^T_\mu(-\mathbf{k})\tau_3(\sigma_j\sigma_2)_{\mu\nu}1_\nu(\mathbf{k})\big)\Big)\\
&+g^{s}_{E_u}\Big(\big(1^\dagger_\alpha(\mathbf{k})\tau_0(\sigma_2)_{\alpha\beta}3^*_\beta(-\mathbf{k})\big)\big(3^T_\mu(-\mathbf{k})\tau_0(\sigma_2)_{\mu\nu}1_\nu(\mathbf{k})\big)\\
&\qquad+\big(1^\dagger_\alpha(\mathbf{k})\tau_3(\sigma_2)_{\alpha\beta}3^*_\beta(-\mathbf{k})\big)\big(3^T_\mu(-\mathbf{k})\tau_3(\sigma_2)_{\mu\nu}1_\nu(\mathbf{k})\big)\Big)\\
&+g^{t}_{E_g}\Big(\big(1^\dagger_\alpha(\mathbf{k})\tau_1(\sigma_2\sigma_j)_{\alpha\beta}3^*_\beta(-\mathbf{k})\big)\big(3^T_\mu(-\mathbf{k})\tau_1(\sigma_j\sigma_2)_{\mu\nu}1_\nu(\mathbf{k})\big)\\
&\qquad+\big(1^\dagger_\alpha(\mathbf{k})\tau_2(\sigma_2\sigma_j)_{\alpha\beta}3^*_\beta(-\mathbf{k})\big)\big(3^T_\mu(-\mathbf{k})\tau_2(\sigma_j\sigma_2)_{\mu\nu}1_\nu(\mathbf{k})\big)\Big)\\
&+g^{s}_{E_g}\Big(\big(1^\dagger_\alpha(\mathbf{k})\tau_1(\sigma_2)_{\alpha\beta}3^*_\beta(-\mathbf{k})\big)\big(3^T_\mu(-\mathbf{k})\tau_1(\sigma_2)_{\mu\nu}1_\nu(\mathbf{k})\big)\\
&\qquad+\big(1^\dagger_\alpha(\mathbf{k})\tau_2(\sigma_2)_{\alpha\beta}3^*_\beta(-\mathbf{k})\big)\big(3^T_\mu(-\mathbf{k})\tau_2(\sigma_2)_{\mu\nu}1_\nu(\mathbf{k})\big)\Big)\\
&+g^{(13)}_{A_{1g}}\big(1^\dagger_\alpha(\mathbf{k})\tau_0 1^*_\beta(-\mathbf{k})3^T_\beta(-\mathbf{k})\tau_0 3_\alpha(\mathbf{k})\\
&\qquad+3^\dagger_\alpha(\mathbf{k})\tau_0 3^*_\beta(-\mathbf{k})1^T_\beta(-\mathbf{k})\tau_0 1_\alpha(\mathbf{k})\big)\\
&+g^{(13)}_{B_{2g}}\big(1^\dagger_\alpha(\mathbf{k})\tau_3 1^*_\beta(-\mathbf{k})3^T_\beta(-\mathbf{k})\tau_3 3_\alpha(\mathbf{k})\\
&\qquad+3^\dagger_\alpha(\mathbf{k})\tau_3 3^*_\beta(-\mathbf{k})1^T_\beta(-\mathbf{k})\tau_3 1_\alpha(\mathbf{k})\big)
\end{aligned} \quad (10)$$

These 14 invariant couplings can be given a physical meaning by relating them to the 'Bloch' Kanamori couplings $U_a$, $U'_a$, $J_a$, $J'_a$. Where the couplings are split into couplings which are symmetry related. The on-site Bloch-Kanamori Hamiltonian is Eqn(11), which takes into account the iron mirror symmetries.

$$H_{BK}(\mathbf{R} \equiv \mathbf{R_i} + \boldsymbol{\delta}) =$$
$$\frac{1}{2}\sum_m U_m \sum_{\alpha\beta} d^\dagger_{m\alpha}(\mathbf{R}) d_{m\alpha}(\mathbf{R}) d^\dagger_{m\beta}(\mathbf{R}) d_{m\beta}(\mathbf{R})$$
$$+ \frac{1}{2}\sum_{m\neq m'} U'_{mm'} \sum_{\alpha\beta} d^\dagger_{m\alpha}(\mathbf{R}) d_{m\alpha}(\mathbf{R}) d^\dagger_{m'\beta}(\mathbf{R}) d_{m'\beta}(\mathbf{R})$$
$$+ \frac{1}{2}\sum_{m\neq m'} J_{mm'} \sum_{\alpha\beta} d^\dagger_{m\alpha}(\mathbf{R}) d_{m'\alpha}(\mathbf{R}) d^\dagger_{m'\beta}(\mathbf{R}) d_{m\beta}(\mathbf{R})$$
$$+ \frac{1}{2}\sum_{m\neq m'} J'_{mm'} \sum_{\alpha\beta} d^\dagger_{m\alpha}(\mathbf{R}) d_{m'\alpha}(\mathbf{R}) d^\dagger_{m\beta}(\mathbf{R}) d_{m'\beta}(\mathbf{R}) \tag{11}$$

Using the method discussed in Ref[19], we relate the parameters of the symmetry allowed couplings of Eqn(10) to Bloch-Kanamori couplings of Eqn(11). The results are listed in Table I, and again with their corresponding mean field order parameters in Tables III,IV, and V.

Lastly, it should be noted that we consider only tree-level electron-electron Coulomb-type interactions, owing to the existence of electron-like (and no hole-like) Fermi surfaces in monolayer [7] and intercalated FeSe [16, 17]. In literature it has been shown that attractive interactions can be produced in monolayer FeSe by coupling to nematic-orbital [21] and spin [22] fluctuations, which lead to spin-singlet s and d-wave superconductivity respectively. However, this analysis highlights an important problem for these mechanisms, any attractive interaction in the spin-singlet s-wave ($A_{1g}$) and d-wave ($B_{2g}$) channels must overcome the large intra-orbital Hubbard repulsion $U_a$ in order to stabilize pairing.

TABLE I. Invariant Coupling Constants

| | |
|---|---|
| $g^{(1)}_{A_{1g}}$ | $U_1 + J'_{11}$ |
| $g^{(1)}_{B_{2g}}$ | $U_1 - J'_{11}$ |
| $g^{(1)}_{A_{2u}}$ | $U'_1 + J_{11}$ |
| $g^{(1)}_{B_{1u}}$ | $U'_1 - J_{11}$ |
| $g^{(3)}_{A_{1g}}$ | $U_3 + J'_{33}$ |
| $g^{(3)}_{B_{2g}}$ | $U_3 - J'_{33}$ |
| $g^{(3)}_{B_{2u}}$ | $U'_3 + J_{33}$ |
| $g^{(3)}_{A_{1u}}$ | $U'_3 - J_{33}$ |
| $g^{t}_{E_u}$ | $U'_{1X3X} - J_{1X3X}$ |
| $g^{s}_{E_u}$ | $U'_{1X3X} + J_{1X3X}$ |
| $g^{t}_{E_g}$ | $U'_{1X3Y} - J_{1X3Y}$ |
| $g^{s}_{E_g}$ | $U'_{1X3Y} + J_{1X3Y}$ |
| $g^{(13)}_{A_{1g}}$ | $J'_{1X3X} + J'_{1X3Y}$ |
| $g^{(13)}_{B_{2g}}$ | $J'_{1X3X} - J'_{1X3Y}$ |

## III. MODEL-EXPERIMENT COMPARISON AND FITS

In Sec(II) we discussed the low-energy effective theory derived from the FeSe plane's space group symmetry. The low-energy theory captures four electronically relevant bands emerging from two 2D symmetry representations at the M-point, M1 and M3. The energy of these representations at the M-point are $\epsilon_1$ and $\epsilon_3$; the precise geometry of these bands emerging away from the M-point depends on values of the Luttinger invariants: $a_1$, $a_3$, $m_1$, $m_3$, $v$. We constrain our model by fitting the invariants to bulk FeSe ARPES [12], which clearly displays bands originating at the M-point at $\epsilon_1 = -5\,\text{meV}$ and $\epsilon_3 = -55\,\text{meV}$. Because of the shared structural unit (the FeSe plane), the intercalated and monolayer FeSe systems should have the same Luttinger invariants as Table(II), up to a uniform shift in the binding energies $\epsilon_1$ and $\epsilon_3$ by $\sim -50\,\text{meV}$.

TABLE II. Luttinger Invariants: FeSe
These parameters were acquired by fitting Eqn(4) to bulk ARPES data Ref[12].

| | |
|---|---|
| $a_1$ | 782.512 meV $\AA^2$ |
| $a_3$ | -1400 meV $\AA^2$ |
| $\frac{1}{2m_1}$ | -492.01 meV $\AA^2$ |
| $\frac{1}{2m_3}$ | 1494.14 meV $\AA^2$ |
| $v$ | 224.406 meV $\AA$ |

The presence of inter-band spin-orbit coupling has been shown in other Fe-based superconductors [23]; the effect of which is a splitting of the band degeneracy in the direction of the Fermi surface crossing (XM-direction), and an avoidance of the Fermi surface crossing (see Fig 6). The monolayer FeSe shows a bandstructure comprised of two electron-like Fermi surfaces at the M-point with no mixing of the Fermi surfaces up to a ∼5 meV resolution [7]. This constraint on the size of the inter-band spin-orbit coupling $\lambda$, which directly leads to the Fermi surface avoidance, is further supported by photoemission in bulk FeSe, which shows no resolved avoided crossing [12, 13]. However, ARPES in the bulk [13] further shows an 11 meV band splitting below the Fermi level in the ΓM-direction. This band-splitting cannot be described by an inter-band spin-orbit coupling, which opens a splitting of order $\lambda$ in the direction of the crossing, yet only opens a splitting of order $\lambda^2/|\epsilon_1 - \epsilon_3|$ in the ΓM-direction below the Fermi level. Simply put, an inter-band spin-orbit coupling large enough to open the 11 meV splitting below the Fermi level would violate experiment by producing a large Fermi surface avoidance.

In order to resolve this problem, we introduced the symmetry allowed $k$-linear intra-band spin-orbit coupling $\lambda_z$ (and higher order invariants $p_{z1}$ and $p_{z2}$; see Eqn 3). This term directly splits the band crossing below the Fermi level, but does not mix the two bands that cross the Fermi level; thus supporting an 11 meV splitting (shown in the ΓM-direction in Fig 5) while not avoiding the electron Fermi-surfaces.

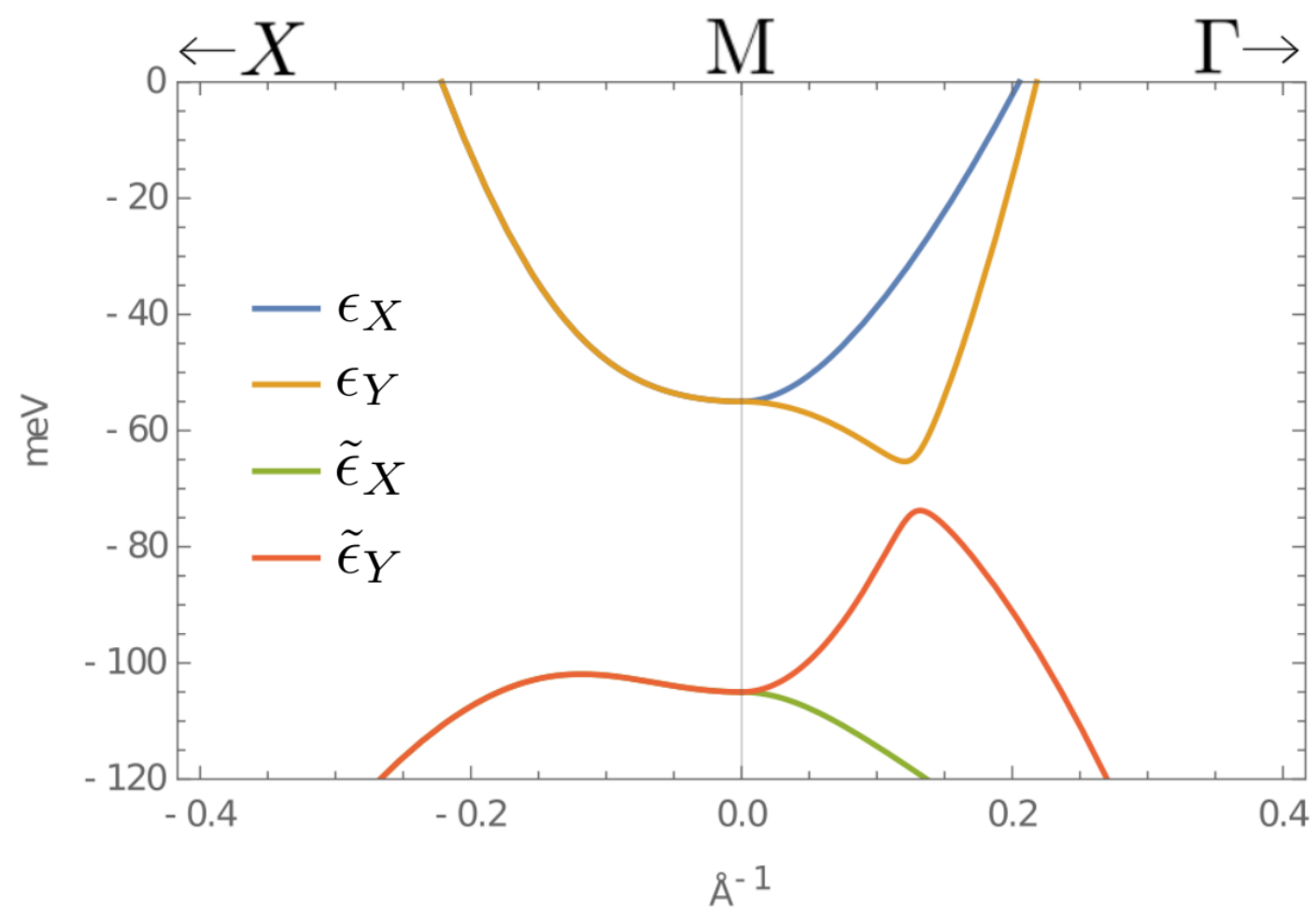

FIG. 5. Bandstructure in the ΓM- and XM-directions. The inter-band spin-orbit coupling is turned off ($\lambda = 0$), thus the bands are four-fold degenerate in the XM-direction. It should be noted that, away from high symmetry directions, each band has a double degeneracy, due to inversion and time-reversal symmetry. The other model parameters used here are the Luttinger invariants Table(II) from fitting Ref[12], and the following: $\lambda_z = 26\,\text{meV}\,\AA^{-1}$, $p_{z1} = p_{z2} = 0$, $\epsilon_1 = -55\,\text{meV}$, and $\epsilon_3 = -105\,\text{meV}$. The bands $\epsilon_X$ and $\tilde{\epsilon}_X$ are the eigenstates of $h'^{\alpha}_X$ (see Eqn 3 and below Eqn 4); the bands $\epsilon_Y$ and $\tilde{\epsilon}_Y$ are the eigenstates of $h'^{\alpha}_Y$, where $\alpha = \uparrow$ or $\downarrow$.

In classifying and disqualifying pairing states, it is practical to constrain the many parameters of the model; however, limiting the study to an over-constrained model risks unjustifiably disqualifying states. Thus we constrain the Luttinger invariants of our model to Table(II), as discussed above, and allow for the values of the bands at the M-point to vary in a $\pm 10\,\text{meV}$ window about $\epsilon_1 = -55\,\text{meV}$ and $\epsilon_3 = -105\,\text{meV}$ (for interclated and monolayer FeSe). With respect to the spin-orbit couplings, we choose $\lambda_z$ (and higher order invariants) as to produce a splitting less than or equal to 20 meV below the Fermi level, and we constrain inter-band spin-orbit $\lambda \leq 5\,\text{meV}$ as discussed previously. These constraints are chosen with the understanding that the key features to reproduce in the normal state are two bands crossing the Fermi level to produce two electron-like Fermi surfaces, and where superconductivity primarily depends on the bandstructure close to the Fermi level.

## IV. PAIRING AND SPIN-ORBIT COUPLING

In the absence of any spin-orbit coupling, the Hamiltonian has full SU(2) spin-symmetry. The intra-band spin-orbit $\lambda_z$ couples the orbital degrees of freedom to the out-of-plane spin, reducing the SU(2) spin-symmetry to U(1). This reduced U(1) symmetry reflects the freedom to rotate the in-plane spin vector $\vec{\sigma} = (\sigma_1, \sigma_2)$ about the z-axis.

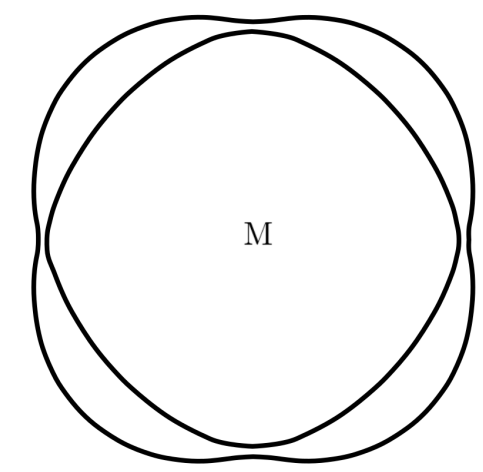

FIG. 6. Sketch of electron pockets at the M-point in the presence of an inter-band spin-orbit coupling $\lambda$. The band degeneracy in the MX-direction is lifted, and the fermisurface crossing becomes avoided. Two new Fermi surfaces are formed – an inner and outer pocket.

In simpler terms, space group operations act on the orbital degrees of freedom ($\tau_i$) while the spin transformations act on the spin degrees of freedom ($\sigma_i$). The introduction of a term in the Hamiltonian that goes as some product of $\tau_i$ and $\sigma_i$ demands that space group and spin transformations transform *together* as to preserve the invariance of the Hamiltonian under space group transformations. The out-of-plane spin-orbit $\lambda_z$ introduces a term in the Hamiltonian that is proportional to $\sigma_z$. Thus while rotations about the z-axis are still symmetries of the Hamiltonian, i.e. $[\sigma_z, H] = 0$, rotations about any in-plane spin axis must be accompanied by rotations in the orbital space.

Only the spin-triplet pairing terms, whose generic form is

$$\vec{d} \cdot \psi^T_{M,\alpha}(-\mathbf{k})\,(i\sigma_2\vec{\sigma})^{\alpha\beta}\,\psi_{M,\beta}(\mathbf{k}),$$

are effected by spin-orbit coupling. Without spin-orbit coupling, spin-triplets have a full SU(2) spin-symmetry and thus the freedom to rotate the $\vec{d}$-vector in any direction. The introduction of $\lambda_z$ decouples the out-of-plane component of $\vec{d}$ from its in-plane component. Triplets formed from in-plane $\vec{d}$-vector, i.e. $i\sigma_2\vec{\sigma} = i\sigma_2(\sigma_1, \sigma_2)$, have their spin-symmetry reduced to U(1), and are free to rotate about the z-axis. On the other hand, the out-of-plane spin-triplet pairs, with $\vec{d}$-vector $i\sigma_2\sigma_3$, become fixed and transform identical to (i.e. same irreducible representation as) a singlet pair.

Before breaking the remaining U(1) spin-symmetry, the in-plane spin-mirror (which is written in terms of Pauli matrices in spin-space as $i\sigma_z$) is a symmetry of the Hamiltonian, thus the orbital and spin degrees of freedom are not constrained to transform together under the in-plane mirror $m_z$. In this scenario spinors $\psi_X$ and $\psi_Y$ transform odd and even under the in-plane mirror respectively, independent of their spin-index. Thus intra-band pairings of the form $\psi_X \otimes \psi_X$ and $\psi_Y \otimes \psi_Y$ transform even under the in-plane mirror; while inter-band pairings of the form $\psi_X \otimes \psi_Y$ transform odd under the in-plane mirror.

TABLE III. Pairing Symmetries with spin-orbit $\lambda_z$ and without spin-orbit $\lambda$. Each pairing is listed with its irreducible symmetry representation and invariant coupling discussed in Sec(II B). (Notation: $\sigma_1 \equiv \sigma_X$, $\sigma_2 \equiv \sigma_Y$, and $\sigma_3 \equiv \sigma_z$.)

| Intra-band $\Delta_X, \Delta_Y$ | | |
|---|---|---|
| Pairing | Symmetry Irrep. | Coupling |
| $1^T \tau_0 i\sigma_2 1$ | $A_{1g}$ | $U_1 + J'_{11}$ |
| $3^T \tau_0 i\sigma_2 3$ | $A_{1g}$ | $U_3 + J'_{33}$ |
| $1^T \tau_3 i\sigma_2 1$ | $B_{2g}$ | $U_1 - J'_{11}$ |
| $3^T \tau_3 i\sigma_2 3$ | $B_{2g}$ | $U_3 - J'_{33}$ |
| $1^T(\tau_0 \pm \tau_3) i\sigma_2 3$ | $E_u$ | $U'_{1X3X} + J_{1X3X}$ |
| $1^T(\tau_0 \pm \tau_3)\sigma_2\sigma_3 3$ | $E_u$ | $U'_{1X3X} - J_{1X3X}$ |
| $1^T(\tau_0 \pm \tau_3)\sigma_2\vec{\sigma} 3$ | $E_u$ x U(1) | $U'_{1X3X} - J_{1X3X}$ |
| Inter-band $\Delta_{XY}, \Delta_{YX}$ | | |
| Pairing | Symmetry Irrep. | Coupling |
| $1^T \tau_1 i\sigma_2 1$ | $A_{2u}$ | $U'_1 + J_{11}$ |
| $3^T \tau_2 i\sigma_2\sigma_3 3$ | $A_{2u}$ | $U'_3 - J_{33}$ |
| $3^T \tau_1 i\sigma_2 3$ | $B_{2u}$ | $U'_3 + J_{33}$ |
| $1^T \tau_2 i\sigma_2\sigma_3 1$ | $B_{2u}$ | $U'_1 - J_{11}$ |
| $1^T \tau_2 i\sigma_2\vec{\sigma} 1$ | $B_{1u}$ x U(1) | $U'_1 - J_{11}$ |
| $3^T \tau_2 i\sigma_2\vec{\sigma} 3$ | $A_{1u}$ x U(1) | $U'_3 - J_{33}$ |
| $1^T(\tau_1 \pm i\tau_2) i\sigma_2 3$ | $E_g$ | $U'_{1X3Y} + J_{1X3Y}$ |
| $1^T(\tau_1 \pm i\tau_2)\sigma_2\sigma_3 3$ | $E_g$ | $U'_{1X3Y} - J_{1X3Y}$ |
| $1^T(\tau_1 \pm i\tau_2)\sigma_2\vec{\sigma} 3$ | $E_g$ x U(1) | $U'_{1X3Y} - J_{1X3Y}$ |

Introducing momentum independent spin-orbit $\lambda$ breaks the remaining spin-symmetry. This means that now $[\sigma_z, H] \neq 0$ and thus there exists no axis of rotation in spin-space about which the Hamiltonian is invariant. The spinors $\psi_X^\alpha$ and $\psi_Y^\alpha$ are now required to transform both orbital and spin together under the in-plane mirror. Consequently, now intra-band pairings with opposite-spin $\psi_X^\uparrow \otimes \psi_X^\downarrow$ and inter-band pairings with same-spin $\psi_X^\uparrow \otimes \psi_Y^\uparrow$ transform even under the in-plane mirror; while intra-band pairings with like-spin $\psi_X^\uparrow \otimes \psi_X^\uparrow$ and inter-band pairings with opposite-spin $\psi_X^\uparrow \otimes \psi_Y^\downarrow$ transform odd.

This in turn divides pairing into two classes: those pairings even under the in-plane mirror Table(IV), and those which are odd Table(V). The even class is represented by the $\Psi_1$ Nambu spinor

$$\Psi_1(\mathbf{k}) = \left(\psi_X^\uparrow(\mathbf{k}), \psi_Y^\downarrow(\mathbf{k}), \psi_X^{\downarrow\,\dagger}(-\mathbf{k}), -\psi_Y^{\uparrow\,\dagger}(-\mathbf{k})\right)^T . \tag{12}$$

The odd class is represented by $\Psi_{2a}$ and $\Psi_{2b}$,

$$\Psi_{2a}(\mathbf{k}) = \left(\psi_X^\uparrow(\mathbf{k}), \psi_Y^\downarrow(\mathbf{k}), \psi_X^{\uparrow\,\dagger}(-\mathbf{k}), -\psi_Y^{\downarrow\,\dagger}(-\mathbf{k})\right)^T ,$$
$$\Psi_{2b}(\mathbf{k}) = \left(\psi_X^\downarrow(\mathbf{k}), \psi_Y^\uparrow(\mathbf{k}), \psi_X^{\downarrow\,\dagger}(-\mathbf{k}), -\psi_Y^{\uparrow\,\dagger}(-\mathbf{k})\right)^T . \tag{13}$$

The two classes are completely independent, with no single irreducible (pairing) representation of the symmetry existing in both classes simultaneously.

The two spinors in the odd class $\Psi_{2a}$ and $\Psi_{2b}$ are related by a spin-flip. They constitute Kramer's pairs, which are independent in the presence of inversion symmetry. This is evident in the double degeneracy of the bands throughout the Brillouin zone. As a consequence of spin-orbit coupling, $\Psi_{2a}$ and $\Psi_{2b}$ map into one another under the $45^\circ$ mirror $m_X$. This mapping of independent spinors into one another allows for odd class pairings states to change sign under the mirror $m_X$ without requiring a node.

TABLE IV. Pairing Symmetries with Spin-Orbit $\lambda$: $\Psi_1$

| Pairing | Symmetry Irrep. | Coupling |
|---|---|---|
| $1^T \tau_0 i\sigma_2 1$ | $A_{1g}$ | $U_1 + J'_{11}$ |
| $3^T \tau_0 i\sigma_2 3$ | $A_{1g}$ | $U_3 + J'_{33}$ |
| $1^T\left((\tau_2 + i\tau_1)i + (\tau_2 - i\tau_1)\sigma_3\right) 3$ | $A_{1g}$ | $U'_{1X3Y} - J_{1X3Y}$ |
| $1^T \tau_3 i\sigma_2 1$ | $B_{2g}$ | $U_1 - J'_{11}$ |
| $3^T \tau_3 i\sigma_2 3$ | $B_{2g}$ | $U_3 - J'_{33}$ |
| $1^T\left((\tau_2 + i\tau_1)(-i) + (\tau_2 - i\tau_1)\sigma_3\right) 3$ | $B_{2g}$ | $U'_{1X3Y} - J_{1X3Y}$ |
| $1^T\left((\tau_2 + i\tau_1)\sigma_3 + (\tau_2 - i\tau_1)(-i)\right) 3$ | $A_{2g}$ | $U'_{1X3Y} - J_{1X3Y}$ |
| $1^T\left((\tau_2 + i\tau_1)\sigma_3 + (\tau_2 - i\tau_1)i\right) 3$ | $B_{1g}$ | $U'_{1X3Y} - J_{1X3Y}$ |
| $1^T(\tau_0 \pm \tau_3) i\sigma_2 3$ | $E_u$ | $U'_{1X3X} + J_{1X3X}$ |
| $1^T(\tau_0 \pm \tau_3)\sigma_2\sigma_3 3$ | $E_u$ | $U'_{1X3X} - J_{1X3X}$ |
| $1^T \tau_2 i\sigma_2\vec{\sigma} 1$ | $E_u$ | $U'_1 - J_{11}$ |
| $3^T \tau_2 i\sigma_2\vec{\sigma} 3$ | $E_u$ | $U'_3 - J_{33}$ |

TABLE V. Pairing Symmetries with Spin-Orbit $\lambda$: $\Psi_{2a}$ ($\Psi_{2b}$)

| Pairing | Symmetry Irrep. | Coupling |
|---|---|---|
| $1^T i\left((\tau_0 + \tau_3)\sigma_3 + (\tau_0 - \tau_3)i\right) 3$ | $A_{1u}$ | $U'_{1X3X} - J_{1X3X}$ |
| $1^T i\left((\tau_0 + \tau_3)i + (\tau_0 - \tau_3)\sigma_3\right) 3$ | $B_{1u}$ | $U'_{1X3X} - J_{1X3X}$ |
| $1^T i\left((\tau_0 + \tau_3)(-i) + (\tau_0 - \tau_3)\sigma_3\right) 3$ | $A_{2u}$ | $U'_{1X3X} - J_{1X3X}$ |
| $1^T \tau_1 i\sigma_2 1$ | $A_{2u}$ | $U'_1 + J_{11}$ |
| $3^T \tau_2 i\sigma_2\sigma_3 3$ | $A_{2u}$ | $U'_3 - J_{33}$ |
| $1^T i\left((\tau_0 + \tau_3)\sigma_3 + (\tau_0 - \tau_3)(-i)\right) 3$ | $B_{2u}$ | $U'_{1X3X} - J_{1X3X}$ |
| $3^T \tau_1 i\sigma_2 3$ | $B_{2u}$ | $U'_3 + J_{33}$ |
| $1^T \tau_2 i\sigma_2\sigma_3 1$ | $B_{2u}$ | $U'_1 - J_{11}$ |
| $1^T(\tau_1 \pm i\tau_2) i\sigma_2 3$ | $E_g$ | $U'_{1X3Y} + J_{1X3Y}$ |
| $1^T(\tau_1 \pm i\tau_2)\sigma_2\sigma_3 3$ | $E_g$ | $U'_{1X3Y} - J_{1X3Y}$ |

### A. Comparison to experiment

The absence of a Fermi surface avoidance (to within $\sim 5\,\mathrm{meV}$ experimental resolution) in the monolayer FeSe

[7] seems to suggest that inter-band spin-orbit coupling $\lambda$ does not play the primary role in superconductivity at the M-point. Although, as we discuss later, it does play a secondary role. In Table(III) we list all symmetry derived pairing states at the M-point in absence of inter-band spin-orbit coupling $\lambda$. They are separated into two symmetry separated classes – intra-band and inter-band pairing. To understand this distinction, note that the energies of the single-particle fermionic excitations are the positive eigenvalues of the matrix

$$\begin{pmatrix} \epsilon_X(\mathbf{k}) & 0 & \Delta_X^* & \Delta_{YX}^* \\ 0 & \epsilon_Y(\mathbf{k}) & \Delta_{XY}^* & \Delta_Y^* \\ \Delta_X & \Delta_{XY} & -\epsilon_X(\mathbf{k}) & 0 \\ \Delta_{YX} & \Delta_Y & 0 & -\epsilon_Y(\mathbf{k}) \end{pmatrix}.$$

The $\epsilon_X(\mathbf{k})$ is the normal state band energy of the pocket-X (i.e. eigenvalue of Eqn 3), and similarly $\epsilon_Y(\mathbf{k})$ is the band energy of the pocket-Y. The intra-band $\Delta_X$ and $\Delta_Y$ pair directly on the Fermi surfaces, and the inter-band $\Delta_{XY}$ and $\Delta_{YX}$ directly pair above/below the Fermi level. From Table(III) it can be seen that there exists no irreducible representation that is simultaneously intra-band and inter-band, and thus no state that will open gaps both on the Fermi level and above/below the Fermi level. As an immediate consequence, all inter-band pairing states are disqualified. This is because inter-band pairing states will not open a gap at the Fermi level (save for at the Fermi surface crossings) for $T = T_c$; thus no superconducting instability exists in those channels.

In this scenario, intra-band pairing is completely independent between the two electron pockets, with no term to mix the two pockets. The problem simplifies into two symmetry related one-band problems. A general property of one-band superconductors is the presence of the gap minimum above the original Fermi surfaces, i.e. back-bending, in agreement with observations in the monolayer [7]. This leaves only the s ($A_{1g}$), d ($B_{2g}$), and p-wave ($E_u$ x U(1)) pairing states in Table(III).

However, STM experiments show a hard gap followed by, not one, but *two* peaks in the $dI/dV$, Fig(3) [6]. The two peaks occur at 8.6 meV and 14.3 meV in intercalated ($Li_{1-x}Fe_x$)OHFeSe [6], and 9 meV and 20.1 meV in monolayer[18]. The suggestion that these two peaks come from the superconducting gap is at odds with monolayer ARPES, whose gap maximum is established to be less than 14 meV (even considering uncertainty) [7]. This suggests the presence of inter-band pairing, where the second peak originates from the gap opened above/below the Fermi level. In fact, without inter-band spin-orbit coupling, there is no way to produce a two-peak $dI/dV$ spectrum. This is because the peaks are indicative of singularities in the density of states [24]. For an anisotropic superconductor these singularities come from the gap maximum, which is a saddle point. Two symmetry related one-band superconductors will each identically produce only one gap and thus only one gap maximum; thus only one peak would be measured by tunnelling.

The simultaneous observation of inter-band phenomena [6, 7, 18] and the experimental constraints on the inter-band spin-orbit ($\lambda \leq 5\,\text{meV}$ [7]), lend themselves towards the existence of a hierarchy. In this hierarchy, the leading order contribution to superconductivity comes from the intra-band pairing states – s, d, and p-wave – and where the introduction of a small inter-band spin-orbit coupling leads to experimentally observed two-peak tunnelling spectrum. The latter is primarily accomplished by the breaking of the SU(2) spin-symmetry, which changes the symmetry of the spin-triplet pairs. This produces symmetry states of both intra- and inter-band character (see Tables IV and V), thus opening gaps above and below the Fermi level. Now a two-peak spectrum is possible, with the second peak coming from the gap opened above the Fermi level (such as is shown in Fig 7).

We studied every possible pure symmetry state and found that only those states which contributed both intra-band and inter-band pairing could generically meet our criteria: (i) nodeless, (ii) two-peak $dI/dV$ spectrum, and (iii) back-bending. Further, because the pairing energy is of the order of half the difference in the band energies, we found that intra-band pairing had to dominate in order to preserve the shape of the back-bending. Scenarios involving only inter-band pairing, or dominated by inter-band pairing, would result in a gap shifted off the normal state Fermi surface, and possibly even merged into a unified gap minimum in between the normal state Fermi surface. Of all symmetry derived pairings at the M-point, only the s, d, and helical p-wave states could meet these criteria.

## V. S-WAVE

In this section we consider time-reversal invariant s-wave superconductivity, both with and without inter-band spin-orbit $\lambda$. Without spin-orbit $\lambda$ the two electron pockets remain independent, and the problem reduces to two independent symmetry related one-band superconductors. Pairing within the electron pocket opens up a gap at the Fermi level; however, no inter-band pairing exists to open a gap above/below the Fermi level. We show that this scenario cannot produce the two-peak tunnelling spectra, as there is no pairing above/below the Fermi level to open a second gap. To complicate matters further, the spin-singlet nature of the s-wave pairing has the added theoretical problem of overcoming the large repulsive intra-orbital Hubbard $U$ (see Table III). Conveniently, the spin-orbit coupling $\lambda$ provides a remedy to both these problems, by introducing an s-wave spin-triplet pair which pairs directly above/below the Fermi level. In this way, a fully gapped spectrum with a two-peak $dI/dV$ and back-bending can be produced. Further, the introduced spin-triplet pair is attractive for (renormalized) Hunds $J$ larger than the inter-orbital Hubbard $U'$, independent of the strength of intra-orbital $U$.

The s-wave symmetry ($A_{1g}$) is invariant under all space group operations. Without the momentum-independent spin-orbit coupling $\lambda$, there are only two $A_{1g}$ pairings:$1^T\Delta_1 i\sigma_2 1$ and $3^T\Delta_3 i\sigma_2 3$. Both pairings are of the intra-band type, constituting two independent symmetry related one-band problems. Choosing to study the X-pocket, we define the Nambu spinor $\Psi_X(\mathbf{k}) = (\psi_X^\uparrow(\mathbf{k}), \psi_X^{\downarrow}{}^\dagger(-\mathbf{k}))^T$. The pairing Hamiltonian is then written as Eqn(14).

$$H_{BdG} = \sum_{\mathbf{k}} \Psi_X^\dagger(\mathbf{k}) \begin{pmatrix} h'^{\uparrow}_X(\mathbf{k}) & \hat{\Delta}_X \\ \hat{\Delta}_X & -h'^{\downarrow T}_X(-\mathbf{k}) \end{pmatrix} \Psi_X(\mathbf{k}) \quad (14)$$

where the $\hat{\Delta}_X$ is the constant 2x2 matrix

$$\hat{\Delta}_X = \begin{pmatrix} \Delta_1 & 0 \\ 0 & \Delta_3 \end{pmatrix}. \quad (15)$$

Noting that because $h'^{\downarrow T}_X(-\mathbf{k}) = h'^{\uparrow}_X(\mathbf{k})$, Eqn(14) can be written as Eqn(16),

$$H_{BdG} = \sum_{\mathbf{k}} \Psi_X^\dagger(\mathbf{k}) \begin{pmatrix} h'^{\uparrow}_X(\mathbf{k}) & \hat{\Delta}_X \\ \hat{\Delta}_X & -h'^{\uparrow}_X(\mathbf{k}) \end{pmatrix} \Psi_X(\mathbf{k}). \quad (16)$$

The normal state Hamiltonian $h'^{\alpha}_X$ has two eigenvalues – one upward and one downward dispersing band. We project $H_{BdG}$ onto the state which crosses the Fermi level, as discussed in Sec(II A).

$$H_{BdG} = \sum_{\mathbf{k}} \Psi_X^\dagger(\mathbf{k}) \begin{pmatrix} \epsilon_X(\mathbf{k}) & \Delta_X(\mathbf{k}) \\ \Delta_X(\mathbf{k}) & -\epsilon_X(\mathbf{k}) \end{pmatrix} \Psi_X(\mathbf{k}) \quad (17)$$

Where the constant 2x2 matrix $\hat{\Delta}_X$ is reduced to a scalar function, inheriting its momentum-dependence from the bandstructure.

$$\begin{aligned} \Delta_X(\mathbf{k}) &= < X\uparrow|\hat{\Delta}_X|X\uparrow > \\ &= \frac{\Delta_1+\Delta_3}{2} + \frac{\Delta_1-\Delta_3}{2}\cos\theta \end{aligned} \quad (18)$$

The pairing function Eqn(18) was previously studied by one of us in a previous work, Ref[19]. In Eqn(17), $\Delta_X(\mathbf{k})$ directly mixes the particle and hole bands, thus opening a gap at the Fermi level. It can be seen from Eqn(18) that the gap anisotropy is centered about the average of the two order parameters $\Delta_1$ and $\Delta_3$, with fluctuations about the average proportional to the difference in the order parameters. If $\Delta_1 = \Delta_3$, the gap is isotropic. For fixed values of the order parameters such that $\Delta_1 \neq \Delta_3$, the fluctuations of the anisotropy about the average depends on $\cos\theta$. Since $\cos\theta$ measures the projection onto the polar axis of the Bloch sphere, which itself represents the reference states M1 and M3, the anisotropies fluctuate largest where the mixing between M1 and M3 is smallest (i.e. the $k_x = -k_y$ direction).

The pairing function for the second electron pocket $\Delta_Y(\mathbf{k})$ is a mirror image of $\Delta_X(\mathbf{k})$ under $m_x$, i.e. $\Delta_X(k_x,k_y) = \Delta_Y(-k_x,k_y)$. No mixing occurs between these two pockets, and no gap is opened above/below the Fermi level.

### A. Pairing with Spin-Orbit Coupling $\lambda$

The inter-band spin-triplet pairing $E_g$ x U(1) directly opens gaps above/below the Fermi level, but does not open any gap at the Fermi level. The momentum-independent spin-orbit coupling $\lambda$ breaks the $E_g$ x U(1) spin-triplet state into four one-dimensional representations, one of which is $A_{1g}$. It is

$$\Delta_t 1^T_\alpha \left( (\tau_2 + i\tau_1) i\delta^{\alpha\beta} + (\tau_2 - i\tau_1)\sigma_3^{\alpha\beta} \right) 3_\beta. \quad (19)$$

Spin-singlets do not change symmetry with spin-orbit coupling, therefore the two original pairings $\Delta_1$ and $\Delta_3$ remain $A_{1g}$. The problem becomes a full 2-band superconductor, with the $H_{BdG}$ written as

$$H_{BdG} = \sum_{\mathbf{k}} \Psi_1^\dagger(\mathbf{k}) \begin{pmatrix} h'^{\uparrow}_X(\mathbf{k}) & \Lambda & \hat{\Delta}_X^\dagger & \hat{\Delta}_{YX}^\dagger \\ \Lambda^\dagger & h'^{\downarrow}_X(\mathbf{k}) & \hat{\Delta}_{XY}^\dagger & \hat{\Delta}_Y^\dagger \\ \hat{\Delta}_X & \hat{\Delta}_{XY} & -h'^{\uparrow}_X(\mathbf{k}) & -\Lambda \\ \hat{\Delta}_{YX} & \hat{\Delta}_Y & -\Lambda^\dagger & -h'^{\downarrow}_X(\mathbf{k}) \end{pmatrix} \Psi_1(\mathbf{k}), \quad (20)$$

where the constant 2x2 pairing matrices are

$$\hat{\Delta}_X = \hat{\Delta}_Y = \begin{pmatrix} \Delta_1 & 0 \\ 0 & \Delta_3 \end{pmatrix} \quad (21)$$

and

$$\hat{\Delta}_{XY} = \hat{\Delta}_{YX}^\dagger = \Delta_t \begin{pmatrix} 0 & i \\ 1 & 0 \end{pmatrix}. \quad (22)$$

Defining the projector

$$\Gamma_1^\dagger = \begin{pmatrix} |X\uparrow> & 0 & 0 & 0 \\ 0 & |Y\downarrow> & 0 & 0 \\ 0 & 0 & |X\uparrow> & 0 \\ 0 & 0 & 0 & |Y\downarrow> \end{pmatrix} \quad (23)$$

allows us to project $H_{BdG}$ onto the reduced band basis, as discussed in Sec(II A). This reduces the size of the Hilbert space by half, greatly simplifying the analysis.

The Hilbert space is halved and the pairing Hamiltonian becomes

$$\mathcal{H}(\mathbf{k}) \equiv \Gamma_1 H_{BdG}(\mathbf{k})\Gamma_1^\dagger = \begin{pmatrix} \epsilon_X(\mathbf{k}) & \lambda\kappa(\mathbf{k}) & \Delta_X(\mathbf{k}) & \Delta_t\kappa(\mathbf{k}) \\ \lambda\kappa^*(\mathbf{k}) & \epsilon_Y(\mathbf{k}) & \Delta_t\kappa^*(\mathbf{k}) & \Delta_Y(\mathbf{k}) \\ \Delta_X(\mathbf{k}) & \Delta_t\kappa(\mathbf{k}) & -\epsilon_X(\mathbf{k}) & -\lambda\kappa(\mathbf{k}) \\ \Delta_t\kappa^*(\mathbf{k}) & \Delta_Y(\mathbf{k}) & -\lambda\kappa^*(\mathbf{k}) & -\epsilon_Y(\mathbf{k}) \end{pmatrix}. \quad (24)$$

The function $\kappa(\mathbf{k})$ is defined as the projection

$$\kappa(\mathbf{k}) = < X\uparrow| \begin{pmatrix} 0 & i \\ 1 & 0 \end{pmatrix} |Y\downarrow> . \quad (25)$$

The phase on both the inter-band mixing $\lambda\kappa$ and the inter-band pairing $\Delta_t\kappa$ both come from $\kappa(k)$. There is no relative phase between the two, thus it is possible to define a unitary transform Eqn(26) which completely eliminates the phase.

$$V_1 = \begin{pmatrix} \frac{\kappa^*}{|\kappa|} & 0 & 0 & 0 \\ 0 & 1 & 0 & 0 \\ 0 & 0 & \frac{\kappa^*}{|\kappa|} & 0 \\ 0 & 0 & 0 & 1 \end{pmatrix} \quad (26)$$

Rotating $\mathcal{H}$ into the new basis via $V_1\mathcal{H}V_1^\dagger$ produces

$$\mathcal{H}(\mathbf{k}) = \begin{pmatrix} \epsilon_X & \lambda|\kappa| & \Delta_X & \Delta_t|\kappa| \\ \lambda|\kappa| & \epsilon_Y & \Delta_t|\kappa| & \Delta_Y \\ \Delta_X & \Delta_t|\kappa| & -\epsilon_X & -\lambda|\kappa| \\ \Delta_t|\kappa| & \Delta_Y & -\lambda|\kappa| & -\epsilon_Y \end{pmatrix}. \quad (27)$$

Let $\tau_i$ and $\sigma_i$ be Pauli-matrices, the 4x4 Hamiltonian Eqn(27) can be written in the convenient form:

$$\mathcal{H} = \tau_3(A + B_3\sigma_3 + B_1\sigma_1) + \tau_1(C + D_3\sigma_3 + D_1\sigma_1), \quad (28)$$

where $A = \frac{1}{2}(\epsilon_X + \epsilon_Y)$, $B_3 = \frac{1}{2}(\epsilon_X - \epsilon_Y)$, $B_1 = \lambda|\kappa|$, $C = \frac{1}{2}(\Delta_X + \Delta_Y)$, $D_3 = \frac{1}{2}(\Delta_X - \Delta_Y)$, and $D_1 = \Delta_t|\kappa|$. And for convenience $B = \sqrt{B_1^2 + B_3^2}$ and $D = \sqrt{D_1^2 + D_3^2}$. The eigenvalues of Eqn(28) is the superconducting dispersion:

$$E_\pm^2 = A^2 + B^2 + C^2 + D^2 \pm 2\sqrt{(AB_3 + CD_3)^2 + (AB_1 + CD_1)^2 + (B_1D_3 - B_3D_1)^2}. \quad (29)$$

Because of the two-band nature of this superconductor, there are two distinct superconducting bands – one upper and one lower. The "upper" and "lower" band are defined by $E_+(\mathbf{k}) > E_-(\mathbf{k})$ for any momentum $\mathbf{k} \in \mathbb{R}_2$. The superconducting gap is the difference between the lower band $E_-$ and the Fermi level, which occurs due to intra-band pairing. In general a second gap exists between the two superconducting bands, defined by $E_+(\mathbf{k}) - E_-(\mathbf{k})$. The splitting $E_+(\mathbf{k}) - E_-(\mathbf{k})$ depends entirely on inter-band pairing, which can come indirectly from $\Delta_X$ and $\Delta_Y$ through $\lambda$, or directly from $\Delta_t$. In Fig(8) we plot the density of states, which is probed by tunneling experiments. Two peaks occur in the spectrum, each from saddle points on the upper and lower superconducting bands. We further plot the dispersion in the ΓM-direction for the same parameters Fig(7), showing the gap lies above the original Fermi surfaces.

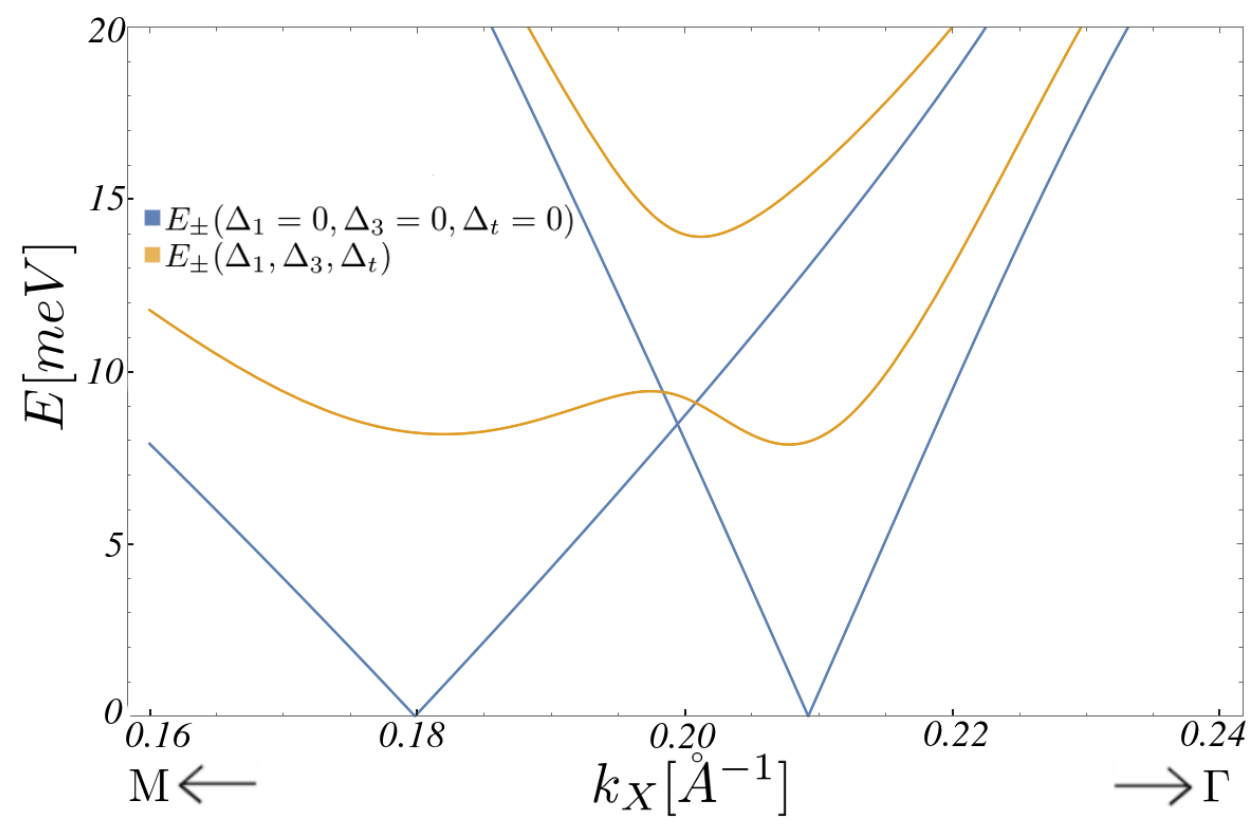

FIG. 7. The upper $E_+$ and lower $E_-$ superconducting band in the ΓM-direction for the $A_{1g}$ state. The blue and yellow curves are the dispersions without and with pairing respectively. In the presence of pairing, the lower band $E_-$ has two local minimum directly above the original Fermi surfaces. The lower and upper bands are split, with the largest contribution to the splitting coming from the inter-band pairing $\Delta_t$. The parameters used are $\Delta_1 = 10.8\,\text{meV}$, $\Delta_3 = 7.2\,\text{meV}$, $\Delta_t = -3\,\text{meV}$, $\lambda = 5\,\text{meV}$, $\epsilon_1 = -45\,\text{meV}$, $\epsilon_3 = -95\,\text{meV}$, $\lambda_z = 26\,\text{meV}\,\AA$, and $p_{z1} = p_{z2} = 0$.

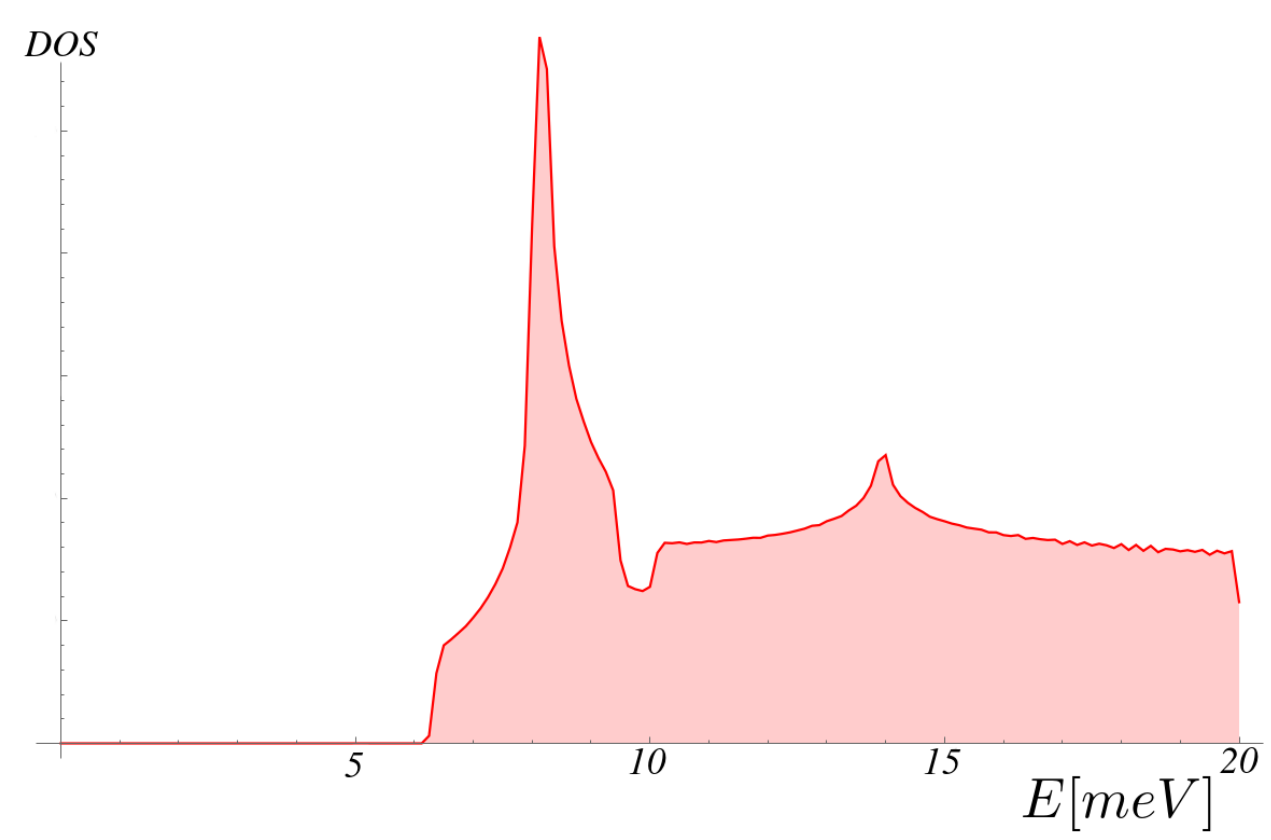

FIG. 8. The Density of States for the $A_{1g}$ state, numerically calculated using the full 8x8 Hamiltonian Eqn(20). Two peaks are present – the first coming from the intra-band pairing gap and the second coming from inter-band pairing above the Fermi level. The few states below the first peak come from the gap minimum, which lies in the direction of the crossing. The same parameters are used as in Fig(7). (*Please note*: a phenomenological scattering model (e.g. Dynes model) was *not* implemented here. Such a model would smooth out the peaks but introduce states into the gap.)

## VI. D-WAVE

The d-wave symmetry is defined by a sign change under a $90^{\circ}$ rotation. There are two crystallographic symmetry representations that have this property: $B_{2g}$ and $B_{1g}$. The $B_{1g}$ representation has one inter-band pairing state, which only appears in the presence of spin-orbit $\lambda$. There are no $B_{1g}$ intra-band pairing states, and thus $B_{1g}$ cannot reproduce the back-bending reported in experiments. On the other hand, the $B_{2g}$ representation has two intra-band pairing states when $\lambda = 0$: $1^T\Delta_1\tau_3 i\sigma_2 1$ and $3^T\Delta_3\tau_3 i\sigma_2 3$. Further, in the presence of $\lambda$, the $B_{2g}$ state picks up a spin-orbit coupled triplet state. This state is an inter-band pairing state, which directly opens a gap above/below the Fermi level. Thus the $B_{2g}$ symmetry representation can produce a two-peak $dI/dV$ spectrum while preserving back-bending.

In this section we consider time-reversal invariant $B_{2g}$ superconductivity in the presence of inter-band spin-orbit $\lambda$. As with s-wave superconductivity, without spin-orbit $\lambda$ the two electron pockets remain independent, and the problem reduces to two independent symmetry related one-band superconductors. Similarly, the spin-singlet d-wave states have the added problem of overcoming the large intra-orbital Hubbard $U$. In fact, without spin-orbit $\lambda$ the d-wave and s-wave problems are identical, save the d-wave superconductor changes sign between pockets. As such, we will not repeat this discussion; we instead point the reader Sec(V).

Two important changes occur with the introduction of spin-orbit $\lambda$. First is the introduction of an inter-band spin-triplet pairing state Eqn(30). Similar to the $A_{1g}$ triplet state discussed in Sec(V A), the $B_{2g}$ triplet pair comes from the reduction of the $E_g$ x U(1) state into one-dimensional representations:

$$\Delta_t 1^T_\alpha \left((\tau_2 + i\tau_1)(-i)\delta^{\alpha\beta} + (\tau_2 - i\tau_1)\sigma_3^{\alpha\beta}\right) 3_\beta. \tag{30}$$

The second important change is to the bandstructure. The spin-orbit coupling $\lambda$ is of the inter-band type, mixing $h'^{\alpha}_X$ and $h'^{\beta}_Y$ bands (for $\alpha \neq \beta$), and avoiding the electron pocket crossing (see Fig 6). Because the $B_{2g}$ state changed sign between the original electron pockets, the mixing of the pockets at the crossing demands symmetry required nodes there. More precisely, two nodes exists at every crossing, one each on the inner and outer pocket. However, because of the large size of the superconducting gap [6, 7, 16] and the experimentally constrained size of $\lambda$ [7], the two nodes merge and open up a gap. Theoretically the merging of Dirac gap nodes has already been studied in the context of hole pockets [25, 26], and in electron pockets of monolayer FeSe [22].

The pairing matrix $H_{BdG}$ is written

$$H_{BdG} = \sum_{\mathbf{k}} \Psi_1^\dagger(\mathbf{k}) \begin{pmatrix} h'^{\uparrow}_X(\mathbf{k}) & \Lambda & \hat{\Delta}^\dagger_X & \hat{\Delta}^\dagger_{YX} \\ \Lambda^\dagger & h'^{\downarrow}_X(\mathbf{k}) & \hat{\Delta}^\dagger_{XY} & \hat{\Delta}^\dagger_Y \\ \hat{\Delta}_X & \hat{\Delta}_{XY} & -h'^{\uparrow}_X(\mathbf{k}) & -\Lambda \\ \hat{\Delta}_{YX} & \hat{\Delta}_Y & -\Lambda^\dagger & -h'^{\downarrow}_X(\mathbf{k}) \end{pmatrix} \Psi_1(\mathbf{k}), \tag{31}$$

where the constant 2x2 pairing matrices are

$$\hat{\Delta}_X = -\hat{\Delta}_Y = \begin{pmatrix} \Delta_1 & 0 \\ 0 & \Delta_3 \end{pmatrix} \tag{32}$$

and

$$\hat{\Delta}_{XY} = \hat{\Delta}^\dagger_{YX} = \Delta_t \begin{pmatrix} 0 & i \\ -1 & 0 \end{pmatrix}. \tag{33}$$

Projecting onto the reduced band basis via the projector Eqn(23) produces:

$$\mathcal{H}(\mathbf{k}) \equiv \Gamma_1 H_{BdG}(\mathbf{k})\Gamma_1^\dagger = \begin{pmatrix} \epsilon_X(\mathbf{k}) & \lambda\kappa(\mathbf{k}) & \Delta_X(\mathbf{k}) & \Delta_t\gamma(\mathbf{k}) \\ \lambda\kappa^*(\mathbf{k}) & \epsilon_Y(\mathbf{k}) & \Delta_t\gamma^*(\mathbf{k}) & \Delta_Y(\mathbf{k}) \\ \Delta_X(\mathbf{k}) & \Delta_t\gamma(\mathbf{k}) & -\epsilon_X(\mathbf{k}) & -\lambda\kappa(\mathbf{k}) \\ \Delta_t\gamma^*(\mathbf{k}) & \Delta_Y(\mathbf{k}) & -\lambda\kappa^*(\mathbf{k}) & -\epsilon_Y(\mathbf{k}) \end{pmatrix}. \tag{34}$$

Again, $\kappa(\mathbf{k})$ is defined as the projection Eqn(25). Further, we defined a second projection:

$$\gamma(\mathbf{k}) = < X\uparrow| \begin{pmatrix} 0 & i \\ -1 & 0 \end{pmatrix} |Y\downarrow>. \tag{35}$$

Noting that in the Bloch sphere coordinates $\gamma(\theta,\phi) = \kappa^*(\theta, \pi-\phi)$.

Important to the symmetry of the superconductor is the relative phase between the inter-band spin-orbit coupling $\lambda\kappa$ and the inter-band pairing $\Delta_t\gamma$. It is the difference in the phase between $\kappa$ and $\gamma$ that the symmetry arises. Applying unitary transform Eqn(26) pushes the phase from the spin-orbit coupling onto the pairing, simplifying this picture.

$$\mathcal{H}(\mathbf{k}) = \begin{pmatrix} \epsilon_X & \lambda|\kappa| & \Delta_X & \Delta_t\frac{\kappa^*}{|\kappa|}\gamma \\ \lambda|\kappa| & \epsilon_Y & \Delta_t\frac{\kappa}{|\kappa|}\gamma^* & \Delta_Y \\ \Delta_X & \Delta_t\frac{\kappa^*}{|\kappa|}\gamma & -\epsilon_X & -\lambda|\kappa| \\ \Delta_t\frac{\kappa}{|\kappa|}\gamma^* & \Delta_Y & -\lambda|\kappa| & -\epsilon_Y \end{pmatrix} \tag{36}$$

If we let $\tau_i$ and $\sigma_i$ be Pauli-matrices, the 4x4 Hamiltonian Eqn(36) can be written in the convenient form:

$$\mathcal{H} = \tau_3(A + B_3\sigma_3 + B_1\sigma_1) + \tau_1(C + D_3\sigma_3 + D_1\sigma_1 + D_2\sigma_2), \tag{37}$$

where $A = \frac{1}{2}(\epsilon_X + \epsilon_Y)$, $B_3 = \frac{1}{2}(\epsilon_X - \epsilon_Y)$, $B_1 = \lambda|\kappa|$, $C = \frac{1}{2}(\Delta_X + \Delta_Y)$, $D_3 = \frac{1}{2}(\Delta_X - \Delta_Y)$, $D_1 = \Delta_t\Re(\frac{\kappa}{|\kappa|}\gamma^*)$, and $D_2 = \Delta_t\Im(\frac{\kappa}{|\kappa|}\gamma^*)$. And for convenience $B = \sqrt{B_1^2 + B_3^2}$

and $D = \sqrt{D_1^2 + D_2^2 + D_3^2}$. The superconducting dispersion is thus:

$$E_\pm^2 = A^2 + B^2 + C^2 + D^2 \pm 2\Big((AB_3 + CD_3 - B_1D_2)^2 + (AB_1 + CD_1 + B_3D_2)^2 + (B_1D_3 - B_3D_1 + CD_2)^2\Big)^{1/2} \quad (38)$$

The d-wave symmetry changes sign under $90^\circ$ rotations. For the $B_{2g}$ representation, this implies the inner and outer pockets change sign in the direction of the avoided Fermi surface crossings. Thus this requires the existence of Dirac gap nodes. However, this is only true so long as the pairing energy is less than half the difference in the band energies. In the direction of the crossing, half the difference in the band energies is equal to $\lambda$. In Fig(9) we plot the superconducting gap in the direction of the crossing. For values of $\Delta_1$, $\Delta_3$, and $\Delta_t$ less than spin-orbit $\lambda$, there exists two Dirac gap nodes. As the pairing parameters increase, the gap above/below the Fermi level grows. This pushes the nodes closer together, annihilating them for $\Delta$'s larger than spin-orbit $\lambda$.

In Fig(11) we plot the density of states showing a two-peak spectrum. The first peak comes from the saddle point in the superconducting gap; while the second peak comes from a saddle point in the upper superconducting band $E_+(\mathbf{k})$. The dispersion in the ΓM-direction is plotted in Fig(10) for the same parameters used to produce the two-peaks, showing the presence of back-bending above the original Fermi surfaces.

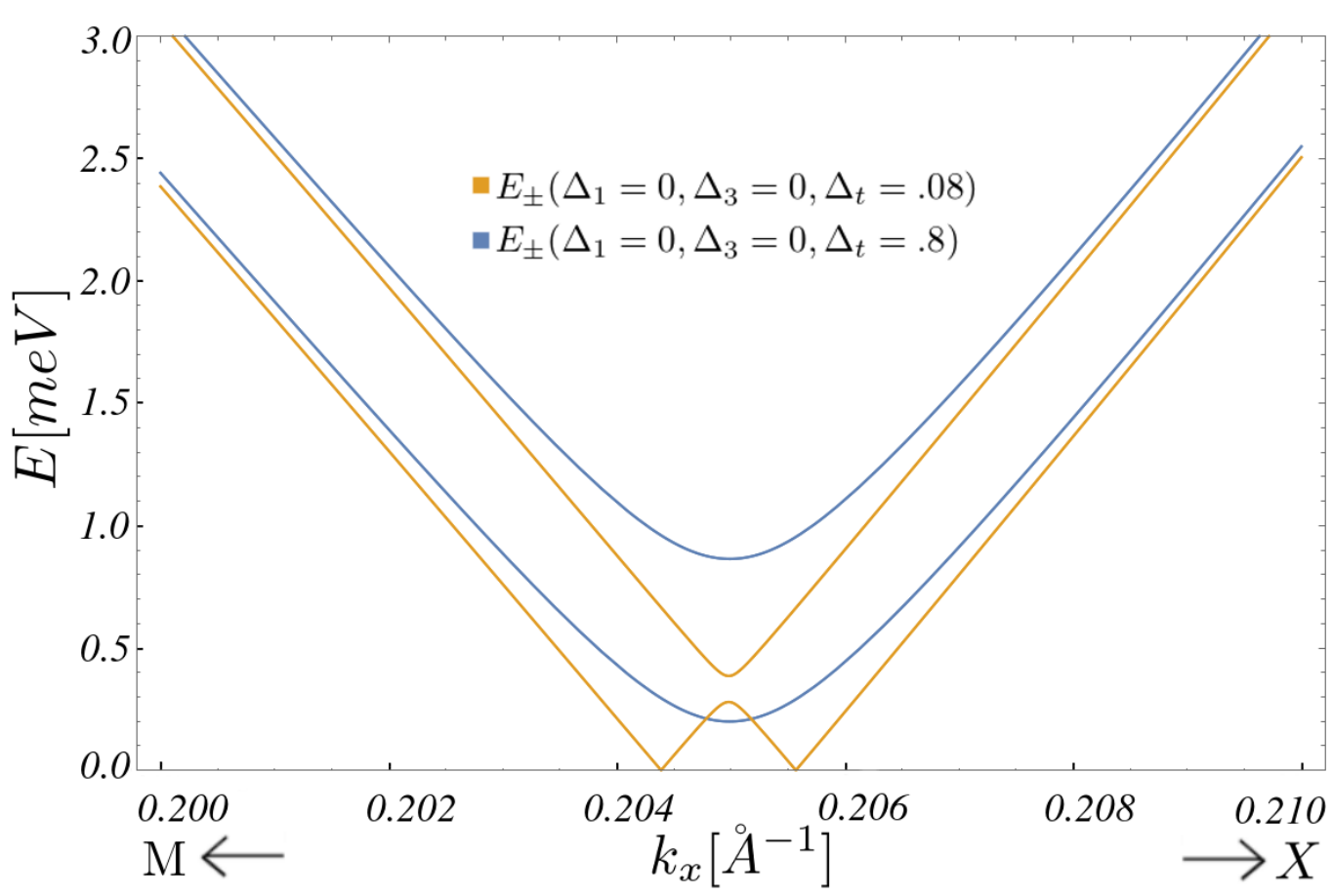

FIG. 9. The superconducting dispersion in the MX-direction for $\Delta_t = .08\,\text{meV}$ (yellow) and $\Delta_t = .8\,\text{meV}$ (blue), and with inter-band spin-orbit $\lambda = .5\,\text{meV}$. For $\Delta < \lambda$, two Dirac gap nodes lie along the $k_x > 0$ line. Inter-band pairing opens a gap above the Fermi level. Increasing $\Delta_t$ grows the gap above the Fermi level, pulling the Dirac nodes together. For $\Delta_t > .5\,\text{meV}$, the nodes merge and open a gap.

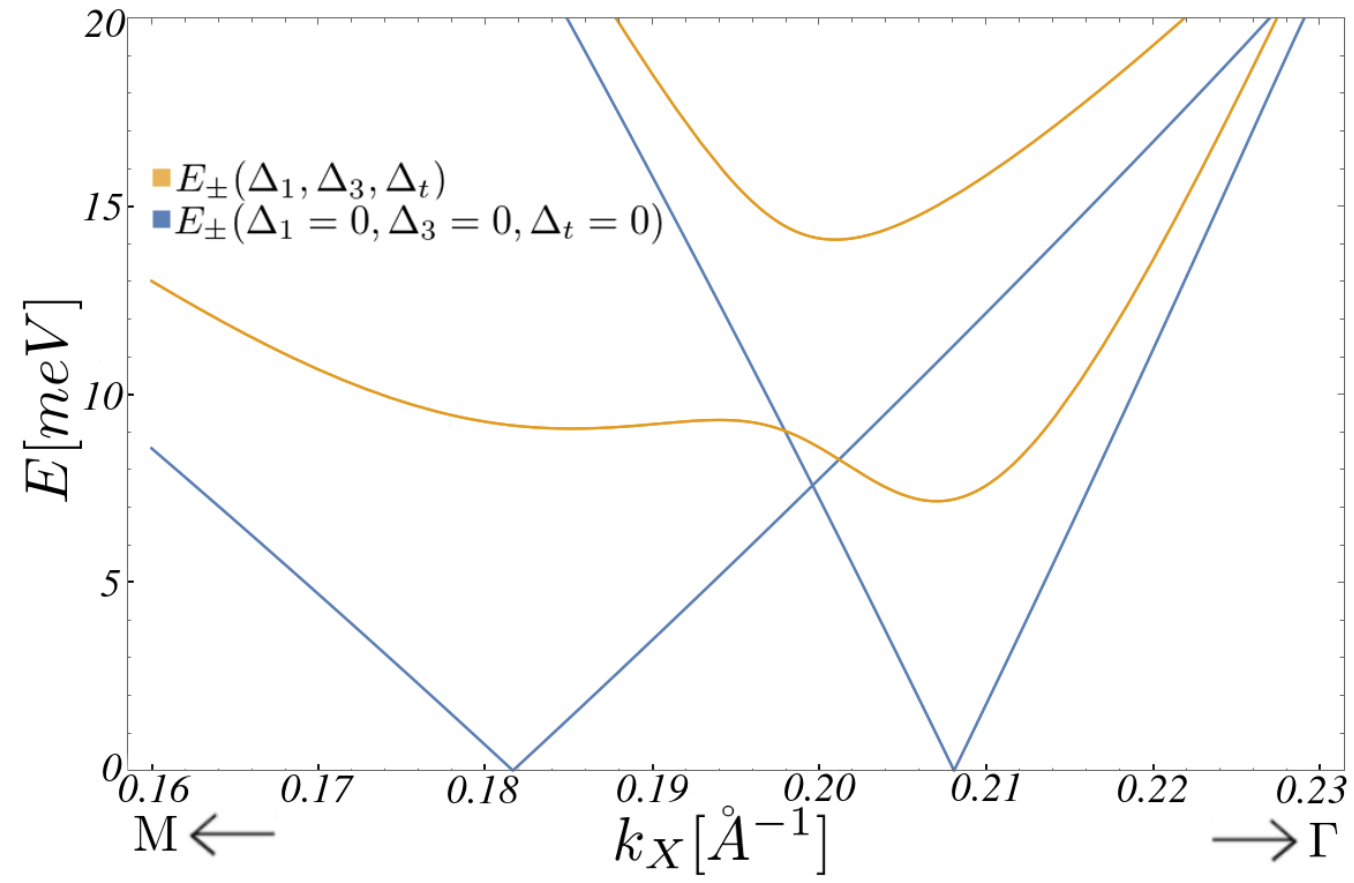

FIG. 10. The upper $E_+$ and lower $E_-$ superconducting band in the ΓM-direction for the $B_{2g}$ state. The blue and yellow curves are the dispersions without and with pairing respectively. In the presence of pairing, the lower band $E_-$ has two local minimum directly above the original Fermi surfaces. The lower and upper bands are split, with the largest contribution to the splitting coming from the inter-band pairing $\Delta_t$. The parameters used are $\Delta_1 = 10.8\,\text{meV}$, $\Delta_3 = 7.2\,\text{meV}$, $\Delta_t = -4.8\,\text{meV}$, $\lambda = .5\,\text{meV}$, $\epsilon_1 = -45\,\text{meV}$, $\epsilon_3 = -95\,\text{meV}$, $\lambda_z = 26\,\text{meV}\,\AA$, and $p_{z1} = p_{z2} = 0$.

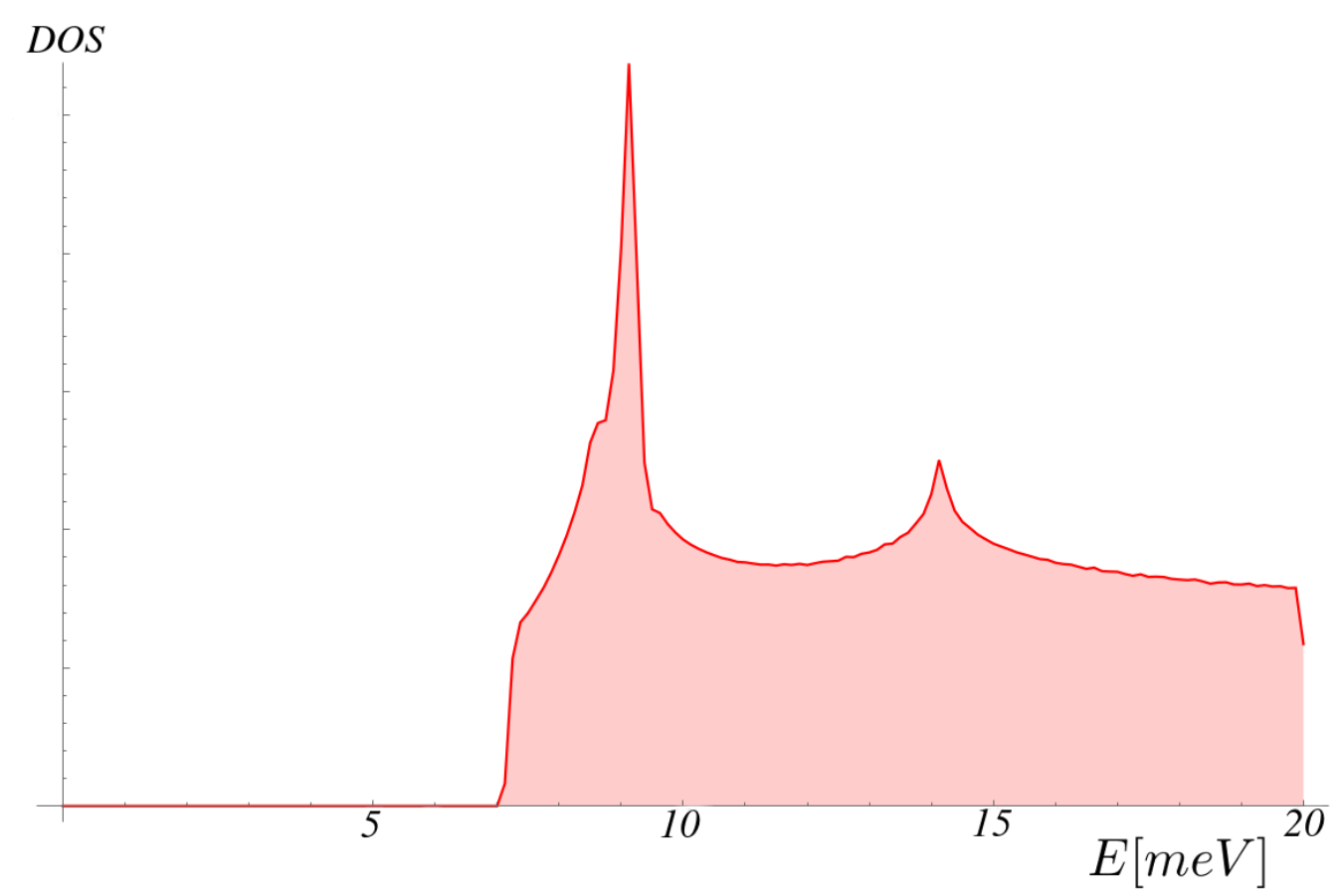

FIG. 11. The Density of States for the $B_{2g}$ state, numerically calculated using the full 8x8 Hamiltonian Eqn(31). Two peaks are present – the first coming from the intra-band pairing gap and the second coming from inter-band pairing above the Fermi level. The same parameters are used as in Fig(10). (*Please note*: a phenomenological scattering model (e.g. Dynes model) was *not* implemented here. Such a model would smooth out the peaks but introduce states into the gap.)

## VII. HELICAL P-WAVE

The p-wave symmetry is characterized by its sign change under inversion. Without inter-band spin-orbit coupling $\lambda$, there exists two p-wave symmetries: $E_u$ and $E_u$ x U(1). The irreducible symmetry representation $E_u$ is also the in-plane polar vector representation $(k_x, k_y)$. The representation labelled purely $E_u$ – composed of one spin-singlet and one spin-triplet with $\vec{d}$-vector pointing out of plane – is nodal. Thus we shift out attention to the second symmetry representation $E_u$ x U(1). This representation is an orbital $E_u$ spin-triplet, with the $\vec{d}$-vector pointing in-plane. The continuous U(1) symmetry represents the freedom to rotate the triplet's $\vec{d}$-vector about the z-axis. This representation

$$1^T_\alpha \Delta_\pm (\tau_0 \pm \tau_3)(\sigma_2 \vec{\sigma})^{\alpha\beta} 3_\beta \tag{39}$$

is a *fully gapped* time-reversal invariant topological superconductor. Its sign change under inversion owes itself to the action of spinors under the generators of symmetry.

Because Eqn(39) is an intra-band pairing, pairing directly on the Fermi level, with no mixing between the independent electron pockets, this problem decouples into two one-band problems. Further, since Eqn(39) pairs electrons with the same spin on the same band, we are forced to define a Nambu spinor with doubled degrees of freedom $\Psi_{X\uparrow}(\mathbf{k}) = (\psi^\uparrow_X(\mathbf{k}), \psi^{\uparrow\dagger}_X(-\mathbf{k}))^T$. This doubling of the degrees of freedom can be corrected by constraining momentum $\mathbf{k}$ to half the Brillouin zone. In this way the spinors $\psi^\uparrow_X(\mathbf{k})$ and $\psi^{\uparrow\dagger}_X(-\mathbf{k})$ are completely decoupled, and the anticommutation relation Eqn(40) is satisfied.

$$\{\psi^\uparrow_X(\mathbf{k}), \psi^{\uparrow\dagger}_X(-\mathbf{k})\} = 0 \tag{40}$$

Because of the presence of time-reversal symmetry, it is not enough to consider only the spin-up particles. A second spinor $\Psi_{X\downarrow}(\mathbf{k}) = (\psi^\downarrow_X(\mathbf{k}), \psi^{\downarrow\dagger}_X(-\mathbf{k}))^T$ exists. Using these spinors, the pairing Hamiltonian is:

$$\begin{aligned} H_{BdG} = \sum_{\mathbf{k}} \Psi^\dagger_{X\uparrow}(\mathbf{k}) \begin{pmatrix} h'^\uparrow_X(\mathbf{k}) & \hat{\Delta}^\dagger_{X\uparrow} \\ \hat{\Delta}_{X\uparrow} & -h'^\uparrow_X(-\mathbf{k})^T \end{pmatrix} \Psi_{X\uparrow}(\mathbf{k}) \\ + \Psi^\dagger_{X\downarrow}(\mathbf{k}) \begin{pmatrix} h'^\downarrow_X(\mathbf{k}) & \hat{\Delta}^\dagger_{X\downarrow} \\ \hat{\Delta}_{X\downarrow} & -h'^\downarrow_X(-\mathbf{k})^T \end{pmatrix} \Psi_{X\downarrow}(\mathbf{k}). \end{aligned} \tag{41}$$

Using the fact that $h'^\uparrow_X(-\mathbf{k})^T = h'^\downarrow_X(\mathbf{k})$, we write:

$$\begin{aligned} H_{BdG} &= H_{BdG\uparrow} + H_{BdG\downarrow} \\ &= \sum_k \Psi^\dagger_{X\uparrow}(\mathbf{k}) \begin{pmatrix} h'^\uparrow_X(\mathbf{k}) & \hat{\Delta}^\dagger_{X\uparrow} \\ \hat{\Delta}_{X\uparrow} & -h'^\downarrow_X(\mathbf{k}) \end{pmatrix} \Psi_{X\uparrow}(\mathbf{k}) \\ &+ \Psi^\dagger_{X\downarrow}(\mathbf{k}) \begin{pmatrix} h'^\downarrow_X(\mathbf{k}) & \hat{\Delta}^\dagger_{X\downarrow} \\ \hat{\Delta}_{X\downarrow} & -h'^\uparrow_X(\mathbf{k}) \end{pmatrix} \Psi_{X\downarrow}(\mathbf{k}). \end{aligned} \tag{42}$$

The U(1) spin-symmetry gives us the freedom to choose any in-plane direction for the $\vec{d}$-vector. We choose to point the $\vec{d}$-vector in the y-direction, writing

$$\hat{\Delta}_{X\uparrow} = \hat{\Delta}_{X\downarrow} = \begin{pmatrix} 0 & \Delta_+ \\ -\Delta_+ & 0 \end{pmatrix}. \tag{43}$$

Under time-reversal spin-up and spin-down particles map into one another as $\psi_{X\uparrow}(\mathbf{k}) \rightarrow \psi_\downarrow(-\mathbf{k})$ and $\psi_{X\downarrow}(\mathbf{k}) \rightarrow -\psi_\uparrow(-\mathbf{k})$; this is followed by a complex conjugation. For this choice of $\vec{d}$-vector $\hat{\Delta}_{X\uparrow} = \hat{\Delta}_{X\downarrow}$ are real matrices. As a consequence, under time-reversal $\psi^T_{X\uparrow}(-\mathbf{k})\hat{\Delta}_{X\uparrow}\psi_{X\uparrow}(\mathbf{k})$ and $\psi^T_{X\downarrow}(-\mathbf{k})\hat{\Delta}_{X\downarrow}\psi_{X\downarrow}(\mathbf{k})$ map into one another exactly. Thus the $E_u$ x U(1) pairing is time-reversally symmetric.

Not only is this pairing time-reversal symmetric, it is fully gapped. In order to show this, it is enough to consider only $H_{BdG\uparrow}$. This is because $H_{BdG\uparrow}$ and $H_{BdG\downarrow}$ are related by time-reversal and share no cross-terms. Defining $\mathcal{H}_\uparrow$ to be the projection of $H_{BdG\uparrow}$ onto the bands that cross the Fermi level, we produce Eqn(44).

$$\mathcal{H}_\uparrow(\mathbf{k}) = \begin{pmatrix} \epsilon_X(\mathbf{k}) & \Delta^*_{X\uparrow}(\mathbf{k}) \\ \Delta_{X\uparrow}(\mathbf{k}) & -\epsilon_X(\mathbf{k}) \end{pmatrix} \tag{44}$$

Projecting the 2x2 constant pairing matrices Eqn(43) into this basis reduces them to scalar functions $\Delta_{X\uparrow}(k)$ and $\Delta_{X\downarrow}(k)$. The momentum dependence of these scalar functions is inherited from the $\epsilon_X(k)$ bandstructure. Inheriting their symmetry from the bandstructure, the up-spin and down-spin pairing functions are related by $\Delta_{X\uparrow}(\theta,\phi) = \Delta_{X\downarrow}(\theta, \pi - \phi)$. Where $\Delta_{X\uparrow}(\mathbf{k})$ in Bloch coordinates $(\theta(\mathbf{k}), \phi(\mathbf{k}))$, is

$$\begin{aligned} \Delta_{X\uparrow}(\mathbf{k}) &\equiv < X\downarrow|\hat{\Delta}_{X\uparrow}|X\uparrow> \\ &= -\sin\theta e^{-i\phi}\Delta_+. \end{aligned} \tag{45}$$

The form factor of Eqn(45) is the azimuthal projection of $\hat{\hbar}$ onto the Bloch sphere. Remembering that in the Bloch representation $\hat{\hbar} = \tau_3 \cos\theta + \sin\theta(\tau_1 \cos\phi + \tau_2 \sin\phi)$. As a consequence of the spin-orbit $\lambda_z$, the projection onto the azimuthal plane of the Bloch sphere is always non-zero. Thus the pairing function $\Delta_{X\uparrow}(\mathbf{k})$ (and by extension $\Delta_{X\downarrow}(\mathbf{k})$) is nodeless.

This state $E_u$ x U(1) is rather novel. While for $\lambda = 0$ the s-wave, d-wave, and p-wave states can produce a full gap at the Fermi level, the p-wave state is the only state which does not have to fight the large intra-orbital repulsion. The spin-triplet nature of the pair provides a pathway for stabilization via the Hunds. The Bloch-Kanamori couplings are listed with their symmetry (for $\lambda = 0$) in Table(III). The coupling for the $E_u$ x U(1) state is $U'_{1X3X} - J_{1X3X}$, which is attractive when the Hunds term $J_{1X3X}$ is greater than the inter-orbital Coulomb repulsion $U'_{1X3X}$.

### A. With spin-orbit $\lambda$

As discussed in Sec(IV), with $\lambda = 0$ the p-wave state is a 1-band problem. This generically leads to gaps opening on the Fermi level but never above/below the Fermi level. Consequently, the $dI/dV$ spectrum will only have one peak at the energy of the gap maximum. Turning on the momentum-independent spin-orbit $\lambda$ resolves this problem. With finite $\lambda$, the full spin-symmetry is broken. This breaks the $E_u$ x U(1) state into four one-dimensional representations, which are listed in Table(V): $A_{1u}$, $B_{1u}$, $A_{2u}$, and $B_{2u}$. Two of these representations – $A_{2u}$ and $B_{2u}$ – contribute two inter-band pairing states which pair directly above/below the Fermi level. The combination of intra-band triplet and inter-band singlets has the ability to produce a two-peak $dI/dV$ spectrum and back-bending.

The $A_{2u}$ and $B_{2u}$ representations are similar, thus it will suffice to discuss the $A_{2u}$ pairing. The three $A_{2u}$ pairings are listed in Table(V). The two inter-band pairings are $1^T\Delta_1\tau_1 i\sigma_2 1$ and $3^T\Delta_3\tau_2 i\sigma_2\sigma_3 3$. The spin-triplet coming from the reduction of $E_u$ x U(1) is:

$$\Delta_t 1^T_\alpha \left((\tau_0+\tau_3)\delta^{\alpha\beta} + (\tau_0-\tau_3) i\sigma_3^{\alpha\beta}\right) 3_\beta. \tag{46}$$

The $A_{2u}$ representation is odd under in-plane mirror $m_z$, thus the pairing Hamiltonian Eqn(47) is written using the $\Psi_{2a}(\mathbf{k})$ and $\Psi_{2b}(\mathbf{k})$ Nambu spinors. As with the $\lambda = 0$ scenario, we avoid the doubling of degrees of freedom by constraining momentum to half the Brillouin zone. This decouples the $\psi_{X\uparrow}(\mathbf{k})$ and $\psi_{X\uparrow}(-\mathbf{k})$ components, satisfying the anticommutation relation Eqn(40).

$$\begin{aligned} H_{BdG} &= H_{BdGa} + H_{BdGb} = \\ &\sum_{\mathbf{k}} \Psi^\dagger_{2a}(\mathbf{k}) \begin{pmatrix} h'^{\uparrow}_X(\mathbf{k}) & \Lambda & \hat{\Delta}^\dagger_X & \hat{\Delta}^\dagger_{YX} \\ \Lambda^\dagger & h'^{\downarrow}_Y(\mathbf{k}) & \hat{\Delta}^\dagger_{XY} & \hat{\Delta}^\dagger_Y \\ \hat{\Delta}_X & \hat{\Delta}_{XY} & -h'^{\downarrow}_X(\mathbf{k}) & \Lambda^* \\ \hat{\Delta}_{YX} & \hat{\Delta}_Y & \Lambda^T & -h'^{\uparrow}_Y(\mathbf{k}) \end{pmatrix} \Psi_{2a}(\mathbf{k}) \\ &+ \Psi^\dagger_{2b}(\mathbf{k}) \begin{pmatrix} h'^{\downarrow}_X(\mathbf{k}) & -\Lambda^* & \hat{\Delta}^\dagger_X & -\hat{\Delta}^T_{YX} \\ -\Lambda^T & h'^{\uparrow}_Y(\mathbf{k}) & -\hat{\Delta}^T_{XY} & -\hat{\Delta}^\dagger_Y \\ \hat{\Delta}_X & -\hat{\Delta}^*_{XY} & -h'^{\uparrow}_X(\mathbf{k}) & -\Lambda \\ -\hat{\Delta}^*_{YX} & -\hat{\Delta}_Y & -\Lambda^\dagger & -h'^{\downarrow}_Y(\mathbf{k}) \end{pmatrix} \Psi_{2b}(\mathbf{k}) \end{aligned} \tag{47}$$

The 2x2 constant intra-band pairing matrices are

$$\hat{\Delta}_X = \Delta_t \begin{pmatrix} 0 & 1 \\ -1 & 0 \end{pmatrix} \tag{48}$$

and

$$\hat{\Delta}_Y = \Delta_t \begin{pmatrix} 0 & i \\ -i & 0 \end{pmatrix}. \tag{49}$$

The 2x2 constant inter-band pairing matrices are

$$\hat{\Delta}_{XY} = \hat{\Delta}_{YX} = \begin{pmatrix} \Delta_1 & 0 \\ 0 & i\Delta_3 \end{pmatrix}. \tag{50}$$

The pairing Hamiltonian $H_{BdGa}$ maps to $H_{BdGb}$ under time-reversal. The two Hamiltonians share no cross terms and are otherwise independent. Thus it is sufficient to focus on $H_{BdGa}$ in studying the superconducting gap structure. Numerical diagonalization of the full 8x8 $H_{BdGa}(\mathbf{k})$ shows two superconducting bands near the Fermi surface, which we refer to as the "upper" $E_+(\mathbf{k})$ and "lower" $E_-(\mathbf{k})$ bands. Plots of both the dispersion in the ΓM-direction Fig(12) and the density of states Fig(13) show both back-bending and a two-peak tunneling spectrum. The two peaks in the spectrum come from saddle points on both the upper and lower superconducting bands. As with the s- and d-wave scenarios, the requirement that there be back-bending implies the dominance of the intra-band pairing. Unlike the s- and d-wave scenarios, however, the dominate intra-band pairing for the p-wave state is a spin-triplet. As mentioned previously, spin-triplet pairs are attractive when the (renormalized) Hunds $J$ overcomes the inter-orbital Hubbard $U'$, independent of the strength of the intra-orbital $U$. For the dominate (i.e. the intra-band) pairing to be spin-triplet supports a hypothesis in which the pairing in this channel is stabilized by the Hunds.

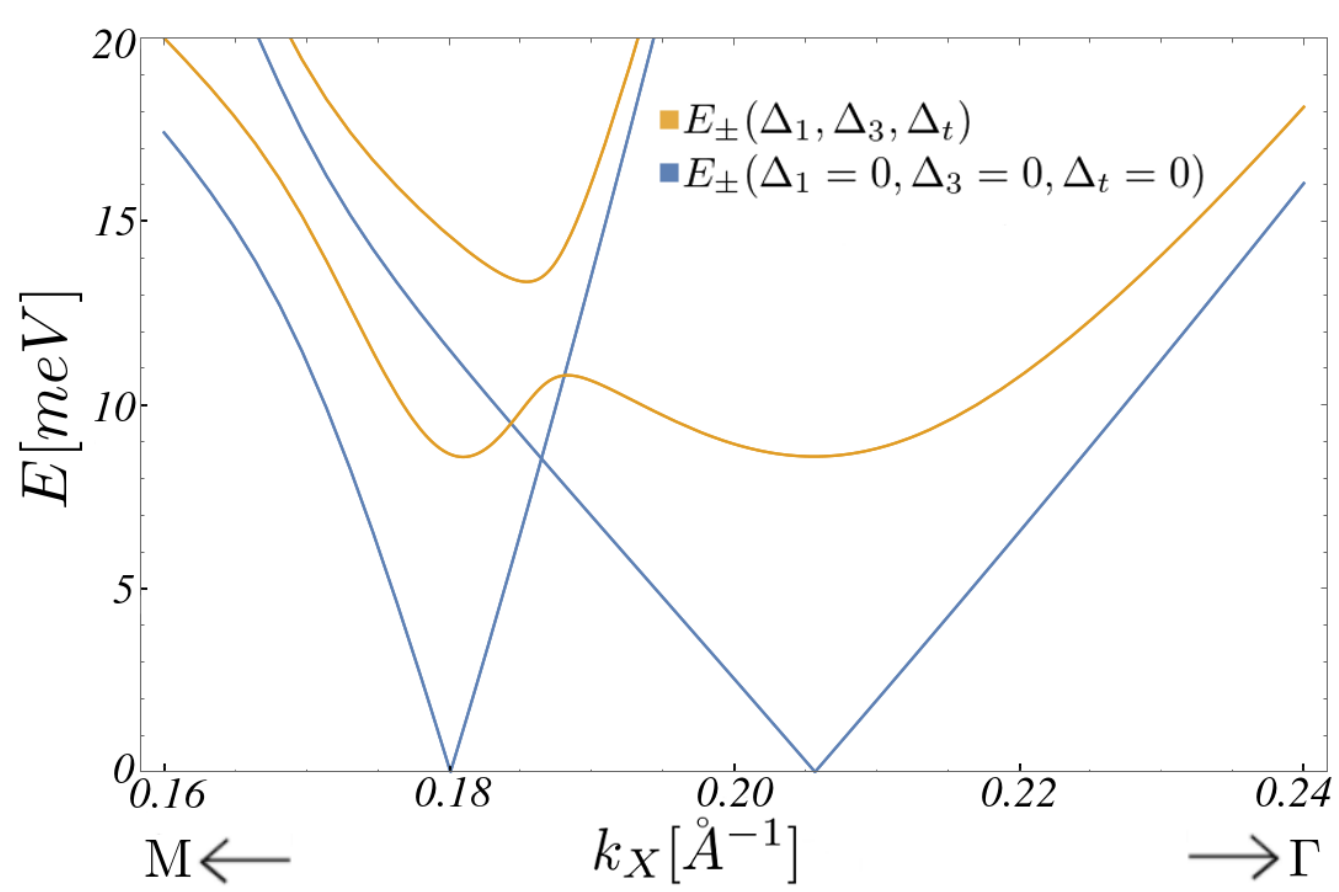

FIG. 12. The upper $E_+$ and lower $E_-$ superconducting band in the ΓM-direction for the $A_{2u}$ state. The blue and yellow curves are the dispersions without and with pairing respectively. In the presence of pairing, the lower band $E_-$ has two local minimum directly above the original Fermi surfaces. The lower and upper bands are split, with the largest contribution to the splitting coming from the inter-band pairing $\Delta_1$ and $\Delta_3$. The parameters used are $\Delta_1 = 6\,\text{meV}$, $\Delta_3 = 7\,\text{meV}$, $\Delta_t = -9\,\text{meV}$, $\lambda = 3\,\text{meV}$, $\epsilon_1 = -55\,\text{meV}$, $\epsilon_3 = -105\,\text{meV}$, $\lambda_z = -31.968\,\text{meV}\,\text{Å}$, $p_{z1} = 4178.88\,\text{meV}\,\text{Å}^3$, and $p_{z2} = 2.5p_{z1}$.

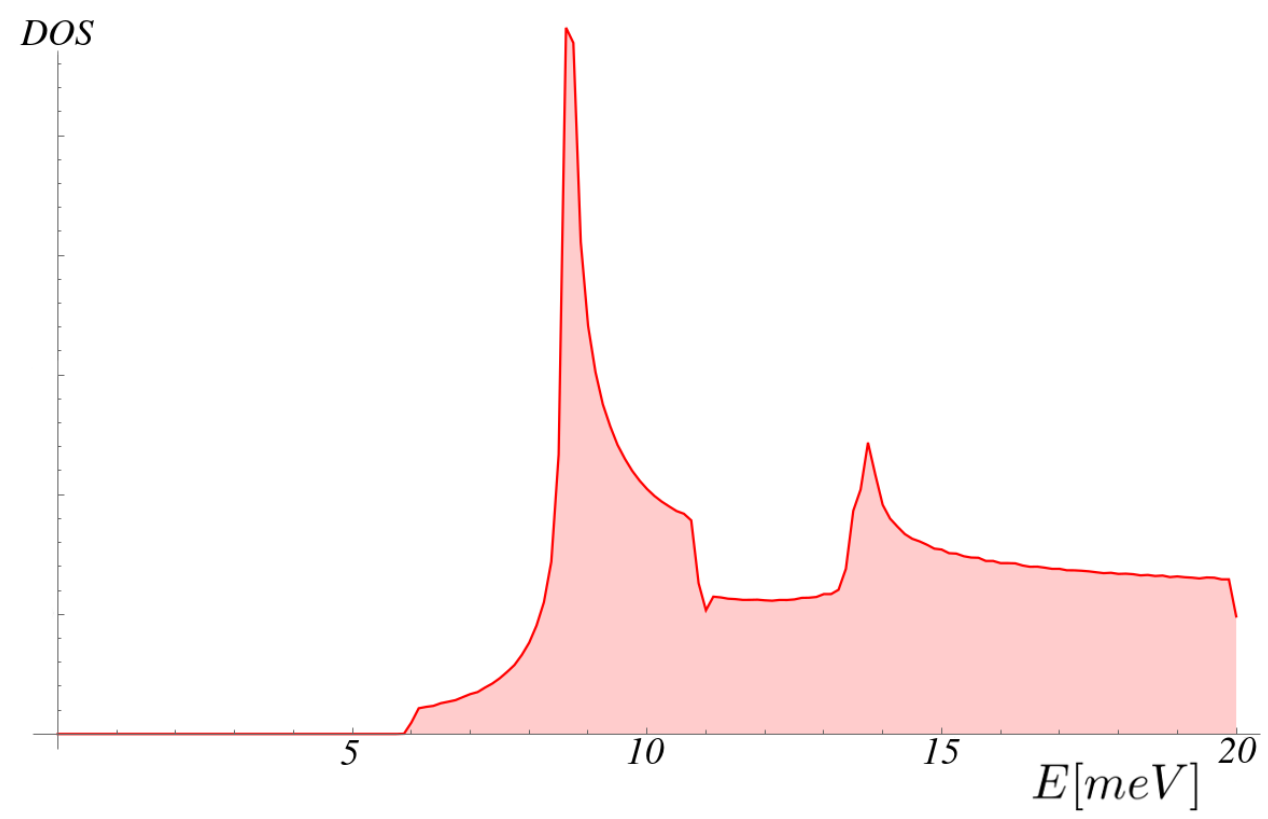

FIG. 13. The Density of States for the $A_{2u}$ state, numerically calculated using the full 8x8 Hamiltonian $H_{BdGa}$ in Eqn(47). Two peaks are present – the first coming from the intra-band pairing gap and the second coming from inter-band pairing above the Fermi level. The few states below the first peak come from the gap minimum, which lies in the direction of the crossing. The same parameters are used as in Fig(12). (*Please note*: a phenomenological scattering model (e.g. Dynes model) was *not* implemented here. Such a model would smooth out the peaks but introduce states into the gap.)

## VIII. CONCLUSION

The constraint on the size of the inter-band mixing ($\lambda \leq 5\,\text{meV}$) from ARPES on the monolayer [7] and bulk [13] suggests the superconducting state and gap in the FeSe's is dominated, to leading order, by pairing *within* two independent electron pockets at M (i.e. intra-band pairing). However, the presence of a two-peak $dI/dV$ STM spectrum [6, 18] in the superconducting state strongly implies the presence of inter-band pairing, which is necessary to open a sizable gap at the band crossing above/below the Fermi level. We studied all symmetry derived order parameters at the M-point in absence of inter-band mixing through the inter-band spin-orbit coupling $\lambda$, and found that no single pairing symmetry simultaneously opens gaps at the Fermi level and above/below the Fermi level (see Table III). Three states open a full gap centered above the original Fermi surfaces – s, d, and helical p-wave intra-band pairing states – but open no gap above/below the Fermi level, and thus cannot reproduce the second tunneling peak [6, 18]. The introduction of a *small* inter-band spin-orbit coupling resolves this problem by breaking the SU(2) spin-symmetry, which changes the symmetry of spin-triplet pairs, and leads to symmetry states of mixed inter- and intra-band pairing character. We further studied all possible pairing symmetries in this scenario (see Table IV and V), and found that only those symmetry states with intra- and inter-band pairing, where the intra-band pairing dominated, was able meet out criteria: (i) full superconducting gap [6, 7, 16, 18], (ii) gap centered above original Fermi surface (i.e. back-bending), and (iii) two-peak Local Density of States (corresponding to a two-peak $dI/dV$ tunnelling spectrum [6, 18]). These states are the s, d, and helical p-wave intra-band pairing states; where inter-band spin-orbit coupling $\lambda$ mixes in some inter-band pairing, leading to a two-peak spectrum as a sub-leading effect.

The connection between the symmetry of the order parameter and the (renormalized) Hubbard-Hunds interactions highlights the detrimental role of the intra- and inter-orbital Hubbard repulsion, $U$ and $U'$, and the beneficial role of the pair-hopping and Hunds interactions, $J'$ and $J$. In particular it reveals that, in the scenario without a small inter-band mixing, the spin-singlet s and d-wave pairing states fight the large intra-orbital Hubbard $U$; while the spin-triplet helical p-wave state benefits from the possibility of an attractive interaction when $U' < J$, completely avoiding the large repulsive intra-orbital $U$ (as was shown for hole pockets in Ref[20]). However, with the introduction of a small inter-band spin-orbit $\lambda$, all three qualifying symmetry states (s,d,and helical p-wave) receive contributions from spin-triplet pairs, and thus all three have an attractive mechanism when $U' < J$, independent of the size of the intra-orbital Hubbard $U$.

We make no conclusion to the exact nature of the superconductivity in FeSe, instead concluding the existence of a pairing hierarchy. This hierarchy is dominated by intra-band pairing, which leads to full gap [6, 7, 16, 18] and back-bending [7]; and where a small inter-band spin-orbit coupling mixes in inter-band pairing, which in turn leads to a two-peak tunnelling spectrum [6, 18] as a sub-leading phenomenon.

We thank J. Kang, A.V. Chubukov, I.I. Mazin, A. Coldea, and R. Fernandes for discussion and input, and a special thanks to M. Watson for his correspondence. OV was supported by NSF DMR-1506756.

## IX. APPENDIX I: GAP MINIMA WITHOUT SPIN-ORBIT COUPLING

ARPES experiment Ref[7] reported no visible Fermi surface avoidance within $\sim 5\,\text{meV}$ resolution. This in turn constrains any inter-band spin-orbit coupling, which would mix the electron-like bands and avoid the Fermi surfaces (see Fig 6). At the same time, STM on the same material (monolayer on $SrTiO_3$) shows the presence of a two features, at energies 9 meV and 20.1 meV [18], with the high energy feature occuring far above the reported gap maximum in Ref[7] (which is less than 14 meV, including uncertainty). In Sec(IV A) we discussed how this, as well as the back-bending [7], implies the existence of a hierarchy, dominated by intra-band pairing along independent electron pockets, and where the second STM feature occurs as a sub-leading phenomenon supported by inter-band pairing. We argued that without an inter-band spin-orbit coupling, there exists no pairing symmetry that opens a gap both at the Fermi level and above/below, which is necessary to produce the two STM features; however, with a small inter-band spin-orbit coupling (constrained to the resolution of the experiment [7]), the SU(2) spin-symmetry is broken, contributing inter-band pairing to the dominate intra-band pairing states.

There is in fact, another piece of evidence for a small sub leading contribution from inter-band pairing. The anisotropy of the superconducting gap in the monolayer as measured by ARPES [7] reports a gap minimum above the original normal state Fermi surface, with the smallest value of the gap occuring above the original Fermi surface crossing. Without inter-band pairing, the electron pockets are independent, and the direction of the crossing constitutes *no special direction*. Why then would the global gap minimum be found there? In fact, without inter-band spin orbit coupling to mix the electron pockets, no global gap minimum (or maximum) will generically occur in that direction. We show that here by a general analysis of pairing in the absence of inter-band spin-orbit coupling, i.e. $\lambda = 0$.

To begin, recognize first that with $\lambda = 0$, the two electron pockets are independent. Thus, because they are related by symmetry, it is sufficient to focus on one electron-like pocket. We choose the X-pocket and define the spin-generalized Nambu spinor $\Psi(k) = (\psi_{X,\alpha}(\mathbf{k}), \psi^{\dagger}_{X,\beta}(-\mathbf{k}))^T$, where $\alpha, \beta \in \{\uparrow, \downarrow\}$. The pairing Hamiltonian is then written as Eqn(51).

$$H_{BdG} = \sum_{\mathbf{k}} \Psi(\mathbf{k})^{\dagger} \begin{pmatrix} h'^{\alpha}_{X}(\mathbf{k}) & \hat{\Delta}^{\dagger}_{X} \\ \hat{\Delta}_{X} & -h'^{\beta}_{X}(-\mathbf{k})^{T} \end{pmatrix} \Psi(\mathbf{k}) \quad (51)$$

The pairing matrix $\hat{\Delta}_X$ is a 2x2 constant matrix; the exact contents of which are fixed by a particular choice of intra-band pairing symmetry in Table(III). For example, the s-wave ($A_{1g}$) pairing state is of the form Eqn(15). We write a generalized pairing matrix as Eqn(52), where $a, b, c, d$ are constants.

$$\hat{\Delta}_X = \begin{pmatrix} a & c \\ d & b \end{pmatrix} \quad (52)$$

Further, notice in Table(III) that there exists no intra-band pairing symmetry that has both diagonal elements of $\hat{\Delta}_X$ and off-diagonal elements of $\hat{\Delta}_X$. This distinction is due to a difference in symmetry between those states of type intra-representational $1 \otimes 1$ and $3 \otimes 3$, and inter-representational $1 \otimes 3$. This produces two possible scenarios: either (i) $a, b \neq 0$ with $c, d = 0$, or (ii) $c, d \neq 0$ with $a, b = 0$.

The 2x2 band Hamiltonian $h'^{\alpha}_{X}$ has two eigenstates, one downward dispersing and one that crosses the Fermi level. We are only interested in the geometry of the gap, which depends on those bands that cross the Fermi level. Projecting onto this reduced band basis simplifies $H_{BdG}$ to a 2x2 matrix, which we define $\mathcal{H}$.

$$\mathcal{H}(\mathbf{k}) = \begin{pmatrix} \epsilon_X(\mathbf{k}) & \Delta^*_X(\mathbf{k}) \\ \Delta_X(\mathbf{k}) & -\epsilon_X(\mathbf{k}) \end{pmatrix} \quad (53)$$

We are interested in the form of the projected pairing function $\Delta_X(\mathbf{k})$, which inherited its momentum dependence from the band basis; defined in Bloch coordinates:

$$\Delta_X(\theta, \phi) = < X, \beta(\theta, \pi - \phi) | \hat{\Delta}_X | X, \alpha(\theta, \phi) > . \quad (54)$$

We now discuss Eqn(54) for all (i) Intra- and (ii) Inter-representational pairing symmetries respectively. We show that the gap anisotropy of (i) takes the form $|\cos\theta|$, and (ii) takes the form of $|\sin\theta|$; neither of which are generically maximum or minimum in the direction of the Fermi surface crossings.

### A. (i) Intra-representational; $a, b \neq 0$, $c, d = 0$

There are two intra-band intra-representational symmetries in Table(III): $A_{1g}$ and $B_{2g}$. Both are spin-singlets, and thus both can be represented by the Nambu spinor $\Psi(\mathbf{k}) = (\psi_{X\uparrow}(\mathbf{k}), \psi_{X\downarrow}(-\mathbf{k}))^T$. The pairing function for these states takes the form Eqn(55).

$$\begin{aligned} \Delta_X(\theta, \phi) &= < X, \uparrow(\theta, \phi) | \hat{\Delta}_X | X, \uparrow(\theta, \phi) > \\ &= \frac{a+b}{2} + \frac{a-b}{2} \cos\theta \end{aligned} \quad (55)$$

For $a, b$ such that $\Delta_X$ is not nodal, the critical point of gap $|\Delta_X|$ occurs where $|\cos\theta|$ is largest.

### B. (ii) Inter-representational; $c, d \neq 0$, $a, b = 0$

There are two intra-band inter-representational symmetries in Table(III): $E_u$ and $E_u$ x U(1). The $E_u$ pairing state is nodal. The $E_u$ x U(1) (helical p-wave) pairing state is a spin-triplet, and for understanding the gap geometry (see Sec VII), is sufficiently represented by the Nambu spinor $\Psi(\mathbf{k}) = (\psi_{X\uparrow}(\mathbf{k}), \psi^{\dagger}_{X\uparrow}(-\mathbf{k}))^T$. The fully gapped pairing function takes the form Eqn(56).

$$\begin{aligned} \Delta_X(\theta, \phi) &= < X, \downarrow(\theta, \phi) | \hat{\Delta}_X | X, \uparrow(\theta, \phi) > \\ &= -(c+d) \cos\frac{\theta}{2} \sin\frac{\theta}{2} e^{-i\phi} \\ &= -\frac{c+d}{2} \sin\theta e^{-i\phi} \end{aligned} \quad (56)$$

Thus the maximum/minimum in the gap $|\Delta_X|$ depends on the maximum/minimum of $|\sin\theta|$.

# X. APPENDIX II: SYMMETRY GENERATORS AND TABLE OF IRREDUCIBLE REPRESENTATIONS

A complete study of the space group symmetry of iron-based superconductors [19] was worked out by Cvetkovic and one of us. In order to assist the reader, we list here all irreducible representations at the Γ-point and those representations at the M-point essential to this paper (i.e. M1 and M3). There are three generators of symmetry (see Fig 1): two mirrors followed by a fractional translation $m_X t$ and $m_z t$, and one mirror $m_x$. With respect to representations of the group $\mathbf{P_\Gamma}$, it is sufficient to consider all three mirrors without fractional translations: $m_X$, $m_z$, and $m_x$. This is because $\mathbf{P_\Gamma}$ is isomorphic to $\mathbf{D_{4h}}$ [19].

TABLE VI. Irreducible Representations of group $\mathbf{P_\Gamma}$. [19]

| $\mathbf{P_\Gamma}$ | $m_X t$ | $m_z t$ | $m_x$ |
|---|---|---|---|
| $A_{1g/u}$ | $\pm1$ | $\pm1$ | $\pm1$ |
| $A_{2g/u}$ | $\mp1$ | $\pm1$ | $\mp1$ |
| $B_{1g/u}$ | $\mp1$ | $\pm1$ | $\pm1$ |
| $B_{2g/u}$ | $\pm1$ | $\pm1$ | $\mp1$ |
| $E_{g/u}$ | $\begin{pmatrix} \pm1 & 0 \\ 0 & \mp1 \end{pmatrix}$ | $\begin{pmatrix} \mp1 & 0 \\ 0 & \mp1 \end{pmatrix}$ | $\begin{pmatrix} 0 & \mp1 \\ \mp1 & 0 \end{pmatrix}$ |

TABLE VII. Irreducible Representations M1 and M3 of group $\mathbf{P_M}$. [19]

| $\mathbf{P_M}$ | $m_X t$ | $m_z t$ | $m_x$ |
|---|---|---|---|
| M1 | $\begin{pmatrix} -1 & 0 \\ 0 & -1 \end{pmatrix}$ | $\begin{pmatrix} -1 & 0 \\ 0 & 1 \end{pmatrix}$ | $\begin{pmatrix} 0 & 1 \\ 1 & 0 \end{pmatrix}$ |
| M3 | $\begin{pmatrix} 1 & 0 \\ 0 & -1 \end{pmatrix}$ | $\begin{pmatrix} -1 & 0 \\ 0 & 1 \end{pmatrix}$ | $\begin{pmatrix} 0 & 1 \\ 1 & 0 \end{pmatrix}$ |